\documentclass[9pt,twocolumn,twoside]{pnas-new}
\usepackage{bm}
\usepackage{pgfplots}
\usepackage{mathtools}
\pgfplotsset{compat=1.14}
\newtheorem{theorem}{Theorem}
\newtheorem{remark}{Remark}
\DeclareMathOperator{\DBS}{DBS}
\DeclareMathOperator{\trace}{trace}
\templatetype{pnasresearcharticle} 

\title{Material Barriers to Diffusive and Stochastic Transport}

\author[a,1]{George Haller}
\author[b]{Daniel Karrasch} 
\author[a]{Florian Kogelbauer}

\affil[a]{Institute for Mechanical Systems, ETH Zürich, Leonhardstrasse 21, 8092 Zürich, Switzerland}
\affil[b]{Zentrum Mathematik, Technische Universität München, Boltzmannstraße 3, 85748 Garching bei München, Germany}

\leadauthor{Haller} 

\significancestatement{
Observations of tracer transport in fluids generally reveal highly complex patterns shaped by an intricate network of transport barriers and enhancers. The elements of this network appear to be universal for small diffusivities, independent of the tracer and its initial distribution. Here, we develop a mathematical theory for weakly diffusive tracers to predict transport barriers and enhancers solely from the flow velocity, without reliance on diffusive or stochastic simulations. Our results yield a simplified computational scheme for diffusive transport problems, such as the estimation of salinity redistribution for climate studies and the forecasting of oil spill spreads on the ocean surface.}

\authorcontributions{G.H. and D.K. designed the research. G.H. developed the theory and wrote the manuscript with contributions from the other authors. D.K. developed the computational algorithm and carried out the numerical simulations. F.K. proved the asymptotic expansion formula for the transport functional.}
\authordeclaration{The authors declare no conflict of interest.}
\correspondingauthor{\textsuperscript{2}To whom correspondence should be addressed. E-mail: georgehaller@ethz.ch}

\keywords{diffusive transport $|$ coherent structures $|$ turbulence $|$ variational calculus $|$} 

\begin{abstract}
We seek transport barriers and transport enhancers as material surfaces
across which the transport of diffusive tracers is minimal or maximal
in a general, unsteady flow. We find that such surfaces are extremizers
of a universal, non-dimensional transport functional whose leading-order
term in the diffusivity can be computed directly from the flow velocity.
The most observable (uniform) transport extremizers are explicitly
computable as null-surfaces of an objective transport tensor. Even in the limit of
vanishing diffusivity, these surfaces differ from  all previously identified coherent
structures for purely advective fluid transport. Our  results extend directly to
stochastic velocity fields and hence enable transport barrier and enhancer
detection under uncertainties.
\end{abstract}

\begin{document}
\verticaladjustment{-2pt}

\maketitle
\thispagestyle{firststyle}
\ifthenelse{\boolean{shortarticle}}{\ifthenelse{\boolean{singlecolumn}}{\abscontentformatted}{\abscontent}}{}

\section{Introduction}

{\it Transport barriers}, i.e., observed inhibitors of the spread of substances in flows, 
provide a simplified global template to analyze mixing without testing various initial concentrations and tracking their pointwise evolution in detail.
Even though such barriers are
well documented in several physical disciplines, including geophysical
flows \cite{weiss08}, fluid dynamics \cite{ottino89}, plasma fusion
\cite{dinklage05}, reactive flows \cite{rosner00} and molecular
dynamics \cite{toda05}, no generally applicable theory for their
defining properties and detection has emerged. In this paper, we seek
to fill this gap by proposing a mathematical theory of transport barriers
and enhancers from first principles in the physically ubiquitous regime
of small diffusivities (high P{\'e}clet numbers). 

\begin{figure}
\centering
\includegraphics[width=1.0\linewidth]{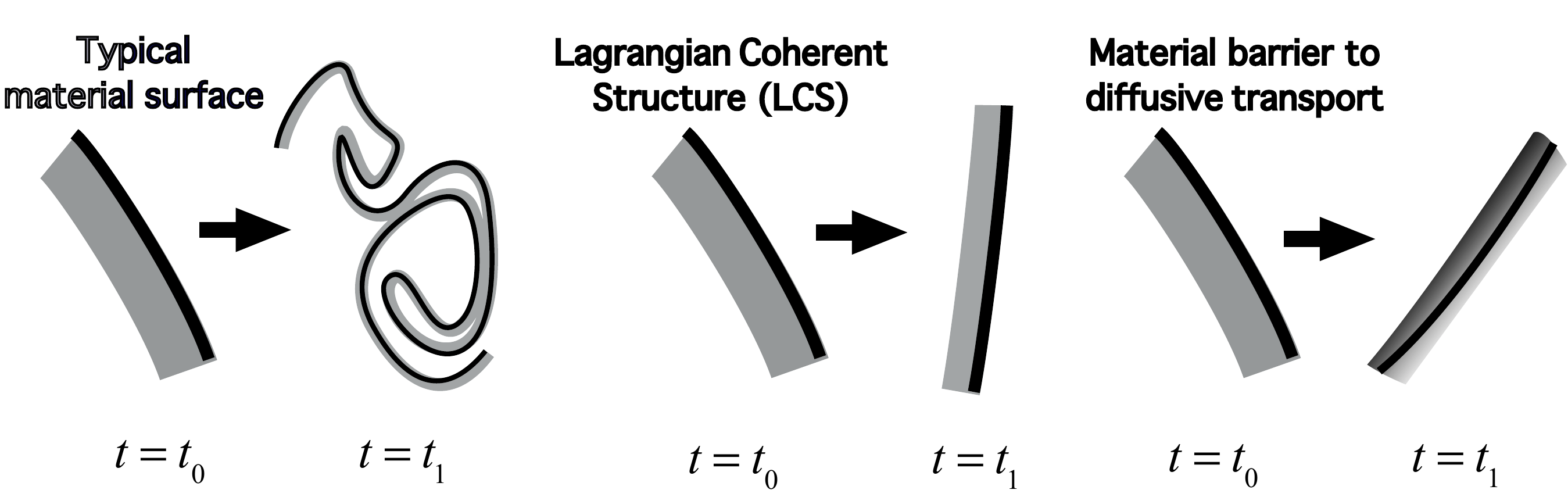}
\caption{Left: Any material surface is a barrier to advective transport over
any time interval $[t_{0},t_{1}]$ but will generally deform into an incoherent shape. Middle: Material surfaces preserving their coherence at their final position at $t_{1}$ are Lagrangian coherent structures (LCSs). Right: Diffusion barriers,
in contrast, are material surfaces minimizing diffusive transport of a concentration field across them.}
\label{fig:transport_barrier}
\end{figure}

Diffusive transport is governed by a time-dependent partial differential
equation (PDE), whose numerical solution requires knowledge of the
initial concentration, the exact diffusivity and the boundary conditions.
Persistently high gradients make this transport PDE challenging to
solve accurately for weakly diffusive processes, such as temperature
and salinity transport in the ocean and vorticity transport in high-Reynolds-number
turbulence. That is why one often neglects diffusion and focuses
on the purely advective redistribution of the substance, governed by
an ordinary differential equation that only involves a deterministic flow velocity
field. In that purely advective setting, a transport barrier is often described
as a surface with zero material flux. While plausible at first sight,
this view actually renders transport barriers grossly ill-defined.
Indeed, \emph{any }codimension-one surface of carrier fluid trajectories
(material surface) experiences zero material flux, and hence is a
barrier by this definition (Fig. \ref{fig:transport_barrier}).

This ambiguity has ignited interest in Lagrangian coherent structures
(LCSs, see Fig. 1), which are material surfaces that do not simply
block but also organize conservative tracers into coherent patterns \cite{peacock10,haller15,bahsoun14,peacock15}.
Due to differing views on finite-time material coherence, however,
each available approach yields (mildly or vastly) different structures
as LCSs \cite{hadjighasem17}. These discrepancies suggest that even purely advective coherent structure detection would benefit from being viewed as the zero-diffusion limit of diffusive barrier detection. Indeed, transport via diffusion through a material surface is a uniquely defined, fundamental physical quantity, whose extremum surfaces can be defined without invoking any special notion of coherence.

A large number of prior approaches to weakly diffusive transport exist,
only some of which will be possible to mention here. Among these,
spatially localized expansions around simple advective solutions provide appealingly detailed
temporal predictions for simple velocity fields \cite{press81,knobloch92,Thiffeault08}. Writing the advection-diffusion
equation in Lagrangian coordinates suggests a quasi-reduction to
a one-dimensional diffusion PDE along the most contracting direction,
yielding asymptotic scaling laws for stretching and folding statistics
along chaotic trajectories \cite{tang96,thiffeault03}. Observed  
transport barriers, however, are not chaotic, and the formal asymptotic
expansions used in these subtle arguments remain
unjustified. As alternatives, the effective diffusivity approach of
\cite{nakamura08} and the residual velocity field concept \cite{pratt16}
offer attractive visualization tools for regions of enhanced or suppressed
transport. Both approaches, however, target already performed diffusive simulations,
and hence provide descriptive diagnostics rather than prediction tools.

Here we address the diffusive tracer transport problem in its purest,
original form. Namely, we seek transport barriers as \emph{space-dividing
(codimension-one) material surfaces that inhibit diffusive transport more than neighboring
surfaces do.} Locating material diffusion barriers without simulating 
diffusion and without reliance on specific initial concentration distributions
is the physical problem we define and solve here in precise mathematical terms, assuming
only incompressibility and small diffusion. In the limit of vanishing
diffusion, our approach also provides a unique, physical definition
of LCSs as material surfaces that will block transport most efficiently under the  
addition of the slightest diffusion or uncertainty to an idealized, purely advective 
mixing problem. Since the notion of transport through a surface is quantitative and universally
accepted, this definition of an LCS eliminates the current ambiguity in advective mixing studies, 
with different approaches identifying different structures as coherent \cite{hadjighasem17}. 

\section{Transport tensor and transport functional}

The advection-diffusion equation for a tracer $c(\mathbf{x},t)$ is
given by \cite{landau66}
\begin{equation}
c_{t}+\mathbf{\bm{\nabla}}\cdot(c\mathbf{v})=\nu\mathbf{\bm{\nabla}}\cdot\left(\mathbf{D}\mathbf{\bm{\nabla}}c\right),\qquad c(\mathbf{x},t_{0})=c_{0}(\mathbf{x}),\label{eq:adv-diff}
\end{equation}
where $\bm{\mathbf{\nabla}}$ denotes the gradient operation with
respect to the spatial variable $\mathbf{x}\in U\subset\mathbb{R}^{n}$
on a compact domain $U$ with $n\geq1$; $\mathbf{v}(\mathbf{x},t)$
is an $n$-dimensional, incompressible, smooth velocity field generating
the advective transport of $c(\mathbf{x},t)$ whose initial distribution
is $c_{0}(\mathbf{x})$; $\mathbf{D}(\mathbf{x},t)=\mathbf{D}^{T}(\mathbf{x},t)\in\mathbb{R}^{n\times n}$
is the dimensionless, positive definite diffusion-structure tensor
describing possible anisotropy and temporal variation in the diffusive
transport of $c$ ; $\nu>0$ is a small diffusivity parameter rendering
the full diffusion tensor $\nu\mathbf{D}$ small in norm. We assume
that the initial concentration $c(\mathbf{x},t_{0})=c_{0}(\mathbf{x})$
is of class $C^{2}$, and the diffusion tensor $\mathbf{D}(\mathbf{x},t)$
is at least H{\"o}lder-continuous, which certainly holds if it is
continuously differentiable. 

The Lagrangian flow map induced by $\mathbf{v}$ is $\mathbf{F}_{t_{0}}^{t}\colon\mathbf{x}_{0}\mapsto\mathbf{x}(t;t_{0},\mathbf{x}_{0})$,
mapping initial material element positions $\mathbf{x}_{0}\in U$ to their later positions
at time $t$. We assume that trajectories stay in the domain $U$ of known velocities, i.e., $\mathbf{F}_{t_{0}}^{t}(U)\subset U$ holds for all
times $t$ of interest. We will denote by $\mathbf{\bm{\nabla}}_{0}\mathbf{F}_{t_{0}}^{t}$
the gradient of $\mathbf{F}_{t_{0}}^{t}$ with respect to initial
positions $\mathbf{x}_{0}$.

Let $\mathcal{M}(t)=\mathbf{F}_{t_{0}}^{t}\left(\mathcal{M}_{0}\right)$
be a time-evolving, $(n-1)$-dimensional material surface in $U$
with boundary $\mathcal{\partial M}(t)$ and with initial position
$\mathcal{M}_{0}=\mathcal{M}(t_{0})$. By construction, the advective
flux of $c$ through $\mathcal{M}(t)$ vanishes and hence only the
diffusive part of the flux vector on the right-hand side of \eqref{eq:adv-diff}
generates transport through $\mathcal{M}(t)$. The total transport
of $c$ through $\mathcal{M}(t)$ over a time interval $[t_{0},t_{1}]$
is therefore given by 
\begin{equation}
\Sigma_{t_{0}}^{t_{1}}=\int_{t_{0}}^{t_{1}}\int_{\mathcal{M}(t)}\nu\mathbf{D}\mathbf{\bm{\nabla}}c\cdot\mathbf{n}\,dA\,dt,\label{eq:transport1}
\end{equation}
with $dA$ denoting the area element on $\mathcal{M}(t)$ and $\mathbf{n}(\mathbf{x},t)$
denoting the unit normal to
$\mathcal{M}(t)$ at a point  $\mathbf{x}\in \mathcal{M}(t)$.  Let $dA_{0}$ and $\mathbf{n}_{0}(\mathbf{x}_{0})$
denote the area element and oriented unit normal vector field on the
initial surface $\mathcal{M}(t_{0})$. Then, by the classic surface
element deformation formula $\mathbf{n}dA=\det\left(\bm{\nabla}_{0}\mathbf{F}_{t_{0}}^{t}\right)\left[\bm{\nabla}_{0}\mathbf{F}_{t_{0}}^{t}\right]^{-\top}\mathbf{n}_{0}dA_{0}$
\cite{gurtin10}, and by the chain rule applied to $\mathbf{\bm{\nabla}}c$,
we can rewrite the total transport \eqref{eq:transport1} through
$\mathcal{M}(t)$ as 
\begin{align}
\Sigma_{t_{0}}^{t_{1}} & =\nu\int_{t_{0}}^{t_{1}}\int_{\mathcal{M}_{0}}\left[\mathbf{\bm{\nabla}}_{0}c\left(\mathbf{F}_{t_{0}}^{t},t\right)\right]^{\top}\mathbf{T}_{t_{0}}^{t}\mathbf{n}_{0}dA_{0}\,dt,\label{eq:transport2}
\end{align}
with the tensor $\mathbf{T}_{t_{0}}^{t}(\mathbf{x}_{0})\in\mathbb{R}^{n\times n}$
defined as
\begin{equation}
\mathbf{T}_{t_{0}}^{t}=\left[\mathbf{\bm{\nabla}}_{0}\mathbf{F}_{t_{0}}^{t}\right]^{-1}\mathbf{D}\left(\mathbf{F}_{t_{0}}^{t},t\right)\left[\mathbf{\bm{\nabla}}_{0}\mathbf{F}_{t_{0}}^{t}\right]^{-\top}.\label{eq:transport tensor}
\end{equation}
 We note that $\det\mathbf{T}_{t_{0}}^{t}=\det\left[\mathbf{D}\left(\mathbf{F}_{t_{0}}^{t},t\right)\right]$
by incompressibility, and that
\begin{equation}
\mathbf{T}_{t_{0}}^{t}=\left[\mathbf{C}_{t_{0}}^{t}\right]^{-1}\label{eq:TCG}
\end{equation}
holds in case of isotropic diffusion ($\mathbf{D}\equiv\mathbf{I}$),
with $\mathbf{C}_{t_{0}}^{t}\coloneqq\left[\mathbf{\bm{\nabla}}_{0}\mathbf{F}_{t_{0}}^{t}\right]^{\top}\mathbf{\bm{\nabla}}_{0}\mathbf{F}_{t_{0}}^{t}$
denoting the Cauchy\textendash Green strain tensor \cite{gurtin10}.

As we show in \emph{SI Appendix S1}, under our assumptions
on $\mathbf{v}$ and $\mathbf{D}$, \eqref{eq:transport2}
can be equivalently re-written as
\begin{equation}
\Sigma_{t_{0}}^{t_{1}}(\mathcal{M}_{0})=\nu\int_{t_{0}}^{t_{1}}\int_{\mathcal{M}_{0}}\left(\mathbf{\bm{\nabla}}_{0}c_{0}\right)^{\top}\mathbf{T}_{t_{0}}^{t}\mathbf{n}_{0}\,dA_{0}\,dt+o(\nu),\label{eq:transport3}
\end{equation}
with the symbol $o(\nu)$ referring to a quantity that, even after
division by $\nu$, tends to zero as $\nu\to0$. Proving \eqref{eq:transport3}
is subtle, because \eqref{eq:adv-diff} is a singularly perturbed PDE for small $\nu>0$, and hence its solutions
generally cannot be Taylor-expanded at $\nu=0$, unless  $\mathbf{v}$ is integrable \cite{liu04}.

To systematically test the ability of the material surface $\mathcal{M}(t)$
to hinder the transport of $c$ over the time interval $\left[t_{0},t_{1}\right]$,
we initialize the concentration field $c$ at time $t_{0}$ locally
near $\mathcal{M}_{0}$ so that $\mathcal{M}_{0}$ is a level surface
of $c_{0}\left(\mathbf{x}_{0}\right)$ along which $\mathbf{\mathbf{\bm{\nabla}}}_{0}c_{0}\left(\mathbf{x}_{0}\right)$
has a constant magnitude $K>0$. This universal choice of $c_{0}\left(\mathbf{x}_{0}\right)$
subjects each $\mathcal{M}_{0}$ surface to the same, most diffusion-prone scalar configuration, ensuring equal detectability for all barriers
in our analysis, independent of any specific initial concentration distribution. We can then write $\bm{\nabla}_{0}c_{0}\left(\mathbf{x}_{0}\right)=K\mathbf{n}_{0}\left(\mathbf{x}_{0}\right)$,
and hence the total transport in \eqref{eq:transport3} becomes
\[
\Sigma_{t_{0}}^{t_{1}}(\mathcal{M}_{0})=\nu K\left(t_{1}-t_{0}\right)\int_{\mathcal{M}_{0}}\left\langle \mathbf{n}_{0},\mathbf{\bar{T}}_{t_{0}}^{t_{1}}\mathbf{n}_{0}\right\rangle dA_{0}+o(\nu).
\]
Here we have introduced the symmetric, positive definite \emph{transport
tensor} $\mathbf{\bar{T}}_{t_{0}}^{t_{1}}$ as the time-average of
$\mathbf{T}_{t_{0}}^{t}$ over $t\in[t_{0},t_{1}].$ The same averaged
tensor was already proposed heuristically in \cite{press81} to simplify
the Lagrangian version of \eqref{eq:adv-diff}.\footnote{ This heuristic simplification generally gives incorrect results for
unsteady flows and can only be partially justified for steady flows
\cite{knobloch92}. In our present context, however, $\mathbf{\bar{T}}_{t_{0}}^{t_{1}}$
arises without any heuristics.}

Finally, to give a dimensionless characterization of the transport
through the surface $\mathcal{M}(t)$ over the period $[t_{0},t_{1}]$,
we normalize $\Sigma_{t_{0}}^{t_{1}}(\mathcal{M}_{0})$ by the diffusivity
$\nu$, by the transport time $\left(t_{1}-t_{0}\right)$, by the
initial concentration gradient magnitude $K$, and by the surface
area $A_{0}(\mathcal{M}_{0})$ (or length, for $n=2$) of $\mathcal{M}_{0}$.
This leads to the normalized total transport
\begin{equation}
\tilde{\Sigma}_{t_{0}}^{t_{1}}(\mathcal{M}_{0}):=\frac{\Sigma_{t_{0}}^{t_{1}}(\mathcal{M}_{0})}{\nu K\left(t_{1}-t_{0}\right)A_{0}(\mathcal{M}_{0})}=\mathcal{T}{}_{t_{0}}^{t_{1}}(\mathcal{M}_{0})+O(\nu^{\alpha})\label{eq:total transport}
\end{equation}
 for some $\alpha\in\left(0,1\right)$, where the non-dimensional \emph{transport functional}
\begin{equation}
\mathcal{T}{}_{t_{0}}^{t_{1}}(\mathcal{M}_{0}):=\frac{\int_{\mathcal{M}_{0}}\left\langle \mathbf{n}_{0},\mathbf{\bar{T}}_{t_{0}}^{t_{1}}\mathbf{n}_{0}\right\rangle dA_{0}}{\int_{\mathcal{M}_{0}}dA_{0}},\label{eq:transport functional}
\end{equation}
is a universal measure of the leading-order diffusive transport through
the material surface $\mathcal{M}(t)$ over the period $[t_{0},t_{1}]$.
This functional enables a systematic comparison of the quality of
transport through different material surfaces. Remarkably, $\mathcal{T}{}_{t_{0}}^{t_{1}}(\mathcal{M}_{0})$
can be computed for any initial surface $\mathcal{M}_{0}$ directly
from the trajectories of \textbf{$\mathbf{v}$}, without solving the
PDE \eqref{eq:adv-diff}. Furthermore, as
we show in \emph{SI Appendix S2}, $\mathbf{\bar{T}}_{t_{0}}^{t_{1}}$
and hence $\mathcal{T}{}_{t_{0}}^{t_{1}}$ are objective (frame-indifferent).

\section{General equation for diffusive transport extremizers}

By formula \eqref{eq:total transport} and by the implicit function
theorem, nondegenerate extrema of the normalized total transport $\tilde{\Sigma}_{t_{0}}^{t_{1}}$
are $O(\nu^{\alpha})$-close to those of the transport
functional $\mathcal{T}{}_{t_{0}}^{t_{1}}$, for some  $\alpha\in\left(0,1\right)$.
Initial positions of such transport-extremizing material surfaces
are, therefore, necessarily solutions of the variational problem
\begin{equation}
\delta\mathcal{T}{}_{t_{0}}^{t_{1}}(\mathcal{M}_{0})=0,\label{eq:main variational problem}
\end{equation}
with boundary conditions yet to be specified, given that the location
and geometry of diffusive transport extremizers is unknown at this
point. We will refer to minimizers of $\mathcal{T}{}_{t_{0}}^{t_{1}}$
as diffusive \emph{transport barriers }and to maximizers of $\mathcal{T}{}_{t_{0}}^{t_{1}}$
as\emph{ }diffusive\emph{ transport enhancers. }

Carrying out the variational differentiation in \eqref{eq:main variational problem}
gives the equivalent extremum problem (cf. \cite{castillo08})
\begin{equation}
\delta\mathcal{E}_{\mathcal{T}_0}\left(\mathcal{M}_{0}\right)=0,\quad\mathcal{E}_{\mathcal{T}_0}\left(\mathcal{M}_{0}\right)\coloneqq\int_{\mathcal{M}_{0}}\left[\left\langle \mathbf{n}_{0},\mathbf{\bar{T}}_{t_{0}}^{t_{1}}\mathbf{n}_{0}\right\rangle -\mathcal{T}_0\right]dA_{0},\label{eq:transport functional 1.5}
\end{equation}
where $\mathcal{T}_0:=\mathcal{T}{}_{t_{0}}^{t_{1}}(\mathcal{M}_{0})$
is constant. To transform this problem to a form amenable to classical
variational calculus, we need to reformulate \eqref{eq:transport functional 1.5}
in terms of a (yet unknown) general parameterization $\mathbf{x}_{0}(s_{1},\ldots,s_{n-1})$
of $\mathcal{M}_{0}$, and then express the integrand in terms of
tangent vectors computed from this parametrization. As we show in
\emph{SI Appendix S3}, if $G_{ij}\left(\partial_{\mathbf{s}}\mathbf{x}_{0}(\mathbf{s})\right)=\left\langle \frac{\partial x_{0}}{\partial s_{i}},\frac{\partial x_{0}}{\partial s_{j}}\right\rangle $,
$i,j=1,\ldots,n-1$ denotes the $(i,j)$ entry of the Gramian matrix
$\mathbf{G}\left(\partial_{\mathbf{s}}\mathbf{x}_{0}(\mathbf{s})\right)$
of the parametrization, then after re-parametrization and passage
from normal to tangent vectors in the integrand, we can rewrite the
functional $\mathcal{E}_{\mathcal{T}_0}$ in \eqref{eq:transport functional 1.5}
in the form
\begin{equation}
\mathcal{E}_{\mathcal{T}_0}\left(\mathcal{M}_{0}\right)=\int_{\mathcal{M}_{0}}L\left(\mathbf{x}_{0}(\mathbf{s}),\partial_{\mathbf{s}}\mathbf{x}_{0}(\mathbf{s})\right)\,ds_{1}\ldots ds_{n-1},\label{eq:transport functional 2}
\end{equation}
with the Lagrangian 
\begin{align}
L\left(\mathbf{x}_{0},\partial_{\mathbf{s}}\mathbf{x}_{0}\right) &=\frac{\det\mathbf{\bar{T}}_{t_{0}}^{t_{1}}\left(\mathbf{x}_{0}\right)\det\left[\mathbf{G}\left(\left(\mathbf{\bar{T}}_{t_{0}}^{t_{1}}\left(\mathbf{x}_{0}\right)\right)^{-\frac{1}{2}}\partial_{\mathbf{s}}\mathbf{x}_{0}\right)\right]}{\sqrt{\det\mathbf{G}\left(\partial_{\mathbf{s}}\mathbf{x}_{0}\right)}}\nonumber \\
 &\phantom{=} -\mathcal{T}_0\sqrt{\det\mathbf{G}\left(\partial_{\mathbf{s}}\mathbf{x}_{0}\right)}.\label{eq:Lagrangian}
\end{align}
The Euler\textendash Lagrange equations associated with the Lagrangian
\eqref{eq:Lagrangian} are given by the $n$-dimensional set of coupled
nonlinear, second-order PDEs
\begin{equation}
\frac{\partial L}{\partial\mathbf{x}_{0}}-\sum_{i=1}^{n-1}\frac{\partial}{\partial s_{i}}\frac{\partial L}{\partial\left(\partial_{s_{i}}\mathbf{x}_{0}\right)}=\mathbf{0}.\label{eq:Euler--Lagrange}
\end{equation}

\section{Uniform extremizers of diffusive transport }

\eqref{eq:Euler--Lagrange} has infinitely many solutions through
any point $\mathbf{x}_{0}$ of the physical space, yet most of these solution surfaces remain unobserved 
as significant barriers due to large variations in the concentration gradient along them. Most observable are 
transport extremizers that maintain a nearly uniform drop in the scalar concentration along them, implying 
that the transport-density along them is as uniform as possible.

As we show in \emph{SI Appendix S4}, even perfectly uniform
extremizers of $\mathcal{T}{}_{t_{0}}^{t_{1}}$ exist and form the
zero level set $\left\{ L=0\right\} $ in the phase space of \eqref{eq:Euler--Lagrange}.
As we see from \eqref{eq:Lagrangian}, these\emph{ uniform transport
extremizer} solutions of \eqref{eq:Euler--Lagrange} satisfy the first-order
family of PDEs
\begin{equation}
\det\mathbf{\bar{T}}_{t_{0}}^{t_{1}}\det\left[\mathbf{G}\left(\left(\mathbf{\bar{T}}_{t_{0}}^{t_{1}}\right)^{-\frac{1}{2}}\partial_{\mathbf{s}}\mathbf{x}_{0}\right)\right]=\mathcal{T}_0\mathbb{\det}\left[\mathbf{G}\left(\partial_{\mathbf{s}}\mathbf{x}_{0}\right)\right],\label{eq:PDE fo runiform extremizers}
\end{equation}
for any choice of the parameter $\mathcal{T}_0>0$. Note that, by construction,
$\mathcal{T}_0$ then equals to the uniform diffusive transport density
across any subset of the material surface $\mathcal{M}(t)$ over the
time interval $[t_{0},t_{1}]$. 

An equivalent form of \eqref{eq:PDE fo runiform extremizers}
follows from the observation that the functional $\mathcal{E}_{\mathcal{T}_0}$
is invariant under reparametrizations and hence $\mathcal{L}_{0}$
can also be computed from the original, surface-normal-based form
\eqref{eq:transport functional 1.5} of the underlying variational
principle. The latter form simply gives $\left\langle \mathbf{n}_{0},\mathbf{\bar{T}}_{t_{0}}^{t_{1}}\mathbf{n}_{0}\right\rangle =\mathcal{T}_0$
on $\mathcal{L}_{0}$, which we further rewrite as
\begin{equation}
\left\langle \mathbf{n}_{0}(\mathbf{x}_{0}),\mathbf{E}_{\mathcal{T}_0}(\mathbf{x}_{0})\mathbf{n}_{0}(\mathbf{x}_{0})\right\rangle =0,\qquad\mathbf{E}_{\mathcal{T}_0}\coloneqq\mathbf{\bar{T}}_{t_{0}}^{t_{1}}-\mathcal{T}_0\mathbf{I}.\label{eq:nullsurface}
\end{equation}
This reveals that diffusive transport extremizers are null-surfaces
of the metric tensor $\mathbf{E}_{\mathcal{T}_0}(\mathbf{x}_{0})$,
i.e., their normals have zero length in the metric defined by $\mathbf{E}_{\mathcal{T}_0}(\mathbf{x}_{0})$. 

For such null-surfaces to exist through a point $\mathbf{x}_{0},$
the metric generated by $\mathbf{E}_{\mathcal{T}_0}$ must have null
directions. This limits the domain of existence of transport extremizers
with uniform transport density $\mathcal{T}_0$ to spatial domains where
the eigenvalues $0<\lambda_{1}(\mathbf{x}_{0})\leq\ldots\leq\lambda_{n}(\mathbf{x}_{0})$
of the positive definite tensor $\mathbf{\bar{T}}_{t_{0}}^{t_{1}}(\mathbf{x}_{0})$
satisfy $\lambda_{1}(\mathbf{x}_{0})\leq\mathcal{T}_0\leq\lambda_{n}(\mathbf{x}_{0})$.

Finding computable sufficient conditions for the solutions of the
variational problem in \eqref{eq:transport functional 1.5} to be
minimizers does not appear to be within reach. Effective necessary
conditions, however, can help greatly in identifying null surfaces
of $\mathbf{E}_{\mathcal{T}_0}(\mathbf{x}_{0})$ that are likely candidates
for extremizers. One such necessary condition requires the trace of
the tensor $\mathbf{E}_{\mathcal{T}_0}$ to be nonnegative, as we show
in \emph{SI Appendix S5}. This enables us to summarize our main results
for transport extremizers in the following theorem.
\begin{theorem}\label{thm1}
A uniform minimizer $\mathcal{M}_{0}$ of the transport functional
$\mathcal{T}{}_{t_{0}}^{t_{1}}$ is necessarily a non-negatively traced
null-surface of the tensor field $\mathbf{E}_{\mathcal{T}_0}$, i.e,
\begin{equation}
\left\langle \mathbf{n}_{0}(\mathbf{x}_{0}),\mathbf{E}_{\mathcal{T}_0}(\mathbf{x}_{0})\mathbf{n}_{0}(\mathbf{x}_{0})\right\rangle =0,\qquad\mathrm{trace\,\mathbf{E}_{\mathcal{T}_0}(\mathbf{x}_{0})\geq0},\label{eq:barriercrit}
\end{equation}
holds at every point $\mathbf{x}_{0}\in\mathcal{M}_{0}$ with unit
normal $\mathbf{n}_{0}(\mathbf{x}_{0})$ to $\mathcal{M}_{0}$. Similarly,
a uniform maximizer $\mathcal{M}_{0}$ of $\mathcal{T}{}_{t_{0}}^{t_{1}}$
is necessarily a non-positively traced null surface of the tensor field
$\mathbf{E}_{\mathcal{T}_0}$, i.e,
\begin{equation}
\left\langle \mathbf{n}_{0}(\mathbf{x}_{0}),\mathbf{E}_{\mathcal{T}_0}(\mathbf{x}_{0})\mathbf{n}_{0}(\mathbf{x}_{0})\right\rangle =0,\qquad\mathrm{trace\,\mathbf{E}_{\mathcal{T}_0}(\mathbf{x}_{0})\leq0},\label{eq:enhancercrit}
\end{equation}
 holds at every point $\mathbf{x}_{0}\in\mathcal{M}_{0}$.\\
\end{theorem}
\begin{remark}
Assume that the flow is two-dimensional ($n=2)$ and the diffusion
is homogeneous and isotropic ($\mathbf{D}=\mathbf{I})$. Then, replacing
the averaged transport tensor $\mathbf{\bar{T}}_{t_{0}}^{t_{1}}$
with its unaveraged counterpart $\mathbf{T}_{t_{0}}^{t_{1}}$ in our
arguments, we obtain that closed material curves that extremize the
diffusive flux uniformly at $t=t_{1}$ coincide with two-dimensional
elliptic Lagrangian coherent structures LCSs \cite{haller13}. Similarly,
replacing $\mathbf{\bar{T}}_{t_{0}}^{t_{1}}$ with the \emph{transport-rate
tensor $\dot{\mathbf{T}}_{t_{0}}^{t_{0}}:=-\left[\bm{\nabla}\mathbf{v}+\left[\bm{\nabla}\mathbf{v}\mathbf{}\right]^{T}\right]$},\footnote{Note that $\dot{\mathbf{T}}_{t_{0}}^{t_{0}}=-2\mathbf{S},$ where
$\mathbf{S}$ is the classic rate-of-strain tensor for the velocity
field $\mathbf{v}.$ } we obtain that closed curves that uniformly extremize the diffusive
flux-rate at $t=t_{0}$ coincide with elliptic objective Eulerian
coherent structures (OECSs) \cite{serra16}.
\end{remark}
Remark 1 connects instantaneous flux and flux-rate extremizing surfaces
under isotropic diffusion to LCSs and EOCSs. In the $\nu\to0$ limit, however, material
diffusion barriers identified by Theorem \ref{thm1} differ from advective coherent
structures identified in previous studies (cf. \emph{SI Appendix S7}). While this conclusion is
at odds with the usual assumptions of purely advective transport studies,
it is mathematically consistent with the singular perturbation nature
of the diffusion term in \eqref{eq:adv-diff}. 
\begin{remark}
As seen in the proof of Theorem \ref{thm1} in \emph{SI Appendix S5}, $\mathrm{trace}\,\mathbf{E}_{\mathcal{T}_0}(\mathbf{x}_{0})=\mathrm{trace}\,\mathbf{\bar{T}}_{t_{0}}^{t_{1}}(\mathbf{x}_{0})-n\mathcal{T}_0$
measures how strongly the normalized transport changes from $\mathcal{T}_0$
under localized normal perturbations at $\mathbf{x}_{0}$ to a transport
extremizer $\mathcal{M}_{0}$. Consequently, the Diffusion Barrier Strength (DBS), defined as
\begin{equation}
\DBS(\mathbf{x}_{0})\coloneqq\trace\mathbf{\bar{T}}_{t_{0}}^{t_{1}}(\mathbf{x}_{0})\label{eq:chidef}
\end{equation}
serves as an objective diagnostic scalar field that highlights centerpieces 
of regions filled with the most influential transport extremizers. Specifically, the time $t_{0}$
positions of the most prevailing diffusion barriers should be marked
approximately by ridges of $\DBS(\mathbf{x}_{0})$ field, while the
time $t_{0}$ positions of the least prevailing diffusion barriers
should be close to trenches of $\DBS(\mathbf{x}_{0})$. A similar conclusion 
holds for diffusion enhancers based on features of the $\DBS(\mathbf{x}_{0})$
field computed in backward time.
\end{remark}
By Remark 2, features of the scalar field $\DBS(\mathbf{x}_{0})$ play a role
analogous to that of the finite-time Lyapunov exponents (FTLEs) in purely
advective transport \cite{haller15}. Unlike the FTLE field, however,
$\DBS(\mathbf{x}_{0})$ is a predictive diagnostic (i.e., requires no diffusive
simulation) and arises directly from the technical construction of diffusion
extremizers (rather than being one possible indicator of their anticipated
properties). Still, $\DBS(\mathbf{x}_{0})$ is a visual diagnostic, while Theorem
\ref{thm1} provides the exact equations that diffusion barriers and enhancers
satisfy.

\section{Application to two-dimensional flows}

Here we solve the general barrier-enhancer equations \eqref{eq:barriercrit}-\eqref{eq:enhancercrit} explicitly for two-dimensional flows and write out a more specific form of the diagnostic $\DBS(\mathbf{x}_{0})$  for such flows. In two dimensions ($n=2)$, a one-dimensional transport extremizer
curve $\mathbf{x}_{0}(s)$ is parametrized by a single scalar parameter
$s\in\mathbb{R}^{1}$. As we show in \emph{SI Appendix S6}, the Lagrangian
$L$ in \eqref{eq:Lagrangian} then simplifies to
\begin{equation}
L(\mathbf{x}_{0},\mathbf{x}_{0}^{\prime})=\frac{\,\left\langle \mathbf{x}_{0}^{\prime},\mathbf{\bar{C}}_{\mathbf{D}}(\mathbf{x}_{0})\mathbf{x}_{0}^{\prime}\right\rangle }{\sqrt{\left\langle \mathbf{x}_{0}^{\prime},\mathbf{x}_{0}^{\prime}\right\rangle }}-\mathcal{T}_0\sqrt{\left\langle \mathbf{x}_{0}^{\prime},\mathbf{x}_{0}^{\prime}\right\rangle },\label{eq:L 2D}
\end{equation}
with the tensor field
\begin{equation}
\mathbf{\bar{C}}_{\mathbf{D}}\coloneqq\frac{1}{t_{1}-t_{0}}\int_{t_{0}}^{t_{1}}\det\left[\mathbf{D}\left(\mathbf{F}_{t_{0}}^{t},t\right)\right]\left[\mathbf{T}_{t_{0}}^{t}\right]^{-1}dt\label{eq:C_D}
\end{equation}
denoting the time-averaged, diffusivity-structure-weighted version
of the classic right Cauchy\textendash Green strain tensor $\mathbf{C}_{t_{0}}^{t}$
introduced in \eqref{eq:TCG}. The Euler\textendash Lagrange
\eqref{eq:Euler--Lagrange} now forms a four-dimensional system of
ODEs\emph{, }which we write out for reference in \emph{SI Appendix
S6}. Uniform transport barriers and enhancers lie in the set $\mathcal{L}_{0}=\left\{ L=0\right\} $
in the $(\mathbf{x}_{0},\mathbf{x}_{0}^{\prime})$ phase space of
this ODE. Equating \eqref{eq:L 2D} with zero, we obtain that solutions
in $\mathcal{L}_{0}$ satisfy $\left\langle \mathbf{x}_{0}^{\prime},\left(\mathbf{\bar{C}}_{\mathbf{D}}(\mathbf{x}_{0})-\mathcal{T}_0\mathbf{I}\right)\mathbf{x}_{0}^{\prime}\right\rangle =0$,
and hence are precisely the null-geodesics of the one-parameter-family
of tensors
\begin{equation}
\hat{\mathbf{E}}_{\mathcal{T}_0}(\mathbf{x}_{0})=\mathbf{\bar{C}}_{\mathbf{D}}(\mathbf{x}_{0})-\mathcal{T}_0\mathbf{I},\label{eq:E tensor}
\end{equation}
which are Lorentzian (i.e., indefinite) metric tensors on the spatial
domain satisfying $\lambda_{1}(\mathbf{x}_{0})<\mathcal{T}_0<\lambda_{2}(\mathbf{x}_{0})$. This extends the mathematical analogy pointed out in \cite{haller13,haller14}
between coherent vortex boundaries and photon spheres around black
holes from advective to diffusive mixing. In this analogy, the role
of the relativistic metric tensor on the four-dimensional space-time
is replaced by the tensor $\mathbf{\hat{E}}_{\mathcal{T}_0}(\mathbf{x}_{0})$
on the two-dimensional physical space of initial conditions. 

We seek unit tangent vectors to null-geodesics of $\hat{\mathbf{E}}_{\mathcal{T}_0}$
as a linear combination $\mathbf{x}_{0}^{\prime}=\bm{\eta}_{\mathcal{T}_0}(\mathbf{x}_{0})=\alpha\mathbf{\bm{\xi}}_{1}\pm\sqrt{1-\alpha^{2}}\mathbf{\bm{\xi}}_{2}$
of the unit eigenvectors $\bm{\xi}_{i}(\mathbf{x}_{0})$ corresponding
to the eigenvalues $0<\lambda_{1}(\mathbf{x}_{0})\leq\lambda_{2}(\mathbf{x}_{0})$
of the positive definite tensor $\mathbf{\bar{C}}_{\mathbf{D}}(\mathbf{x}_{0})$.
Substituting this linear combination into $\left\langle \mathbf{x}_{0}^{\prime},\left(\mathbf{\bar{C}}_{\mathbf{D}}(\mathbf{x}_{0})-\mathcal{T}_0\mathbf{I}\right)\mathbf{x}_{0}^{\prime}\right\rangle =0$
and solving for $\alpha\in[0,1]$ gives the direction field family
\begin{equation}
\mathbf{x}_{0}^{\prime}=\bm{\eta}_{\mathcal{T}_0}(\mathbf{x}_{0})\coloneqq\sqrt{\tfrac{\lambda_{2}-\mathcal{T}_0}{\lambda_{2}-\lambda_{1}}}\mathbf{\bm{\xi}}_{1}\pm\sqrt{\tfrac{\mathcal{T}_0-\lambda_{1}}{\lambda_{2}-\lambda_{1}}}\mathbf{\bm{\xi}}_{2}\label{eq:eta field}
\end{equation}
for null-geodesics of $\hat{\mathbf{E}}_{\mathcal{T}_0}$, defined only
on the domain where $\lambda_{1}(\mathbf{x}_{0})\leq\mathcal{T}_0\leq\lambda_{2}(\mathbf{x}_{0}).$
Trajectories of $\bm{\eta}_{\mathcal{T}_0}$ experience uniform
pointwise transport density $\mathcal{T}_0$ over the time interval
$[t_{0},t_{1}].$ For homogeneous, isotropic diffusion ($\mathbf{D}\equiv\mathbf{I}$),
we have $\mathbf{\bar{T}}_{t_{0}}^{t_{1}}=\mathbf{\bar{C}}_{\mathbf{D}}^{-1}$
by incompressibility (cf.~\emph{SI Appendix S6}). Consequently, the scalar diagnostic featured in Remark 2 takes the specific form $\mathrm{DBS}(\mathbf{x}_{0})=\lambda_{1}(\mathbf{x}_{0})+\lambda_{2}(\mathbf{x}_{0})$.
Finally, as we show in \emph{SI Appendix S6}, there are only three types of robust barriers to diffusion in two-dimensional flows:
fronts, jet cores and families of closed material curves forming material
vortices. This is consistent with observations of large-scale geophysical flows \cite{weiss08}.
\section{Particle transport extremizers in stochastic velocity fields}

Here, we show how our results on barriers to diffusive scalar transport carry over to probabilistic transport barriers to fluid particle motion with uncertainties. Such motions are typically modeled by diffusive It{\^o}
processes of the form 
\begin{equation}
d\mathbf{x}(t)=\mathbf{v}(\mathbf{x}(t),t)dt+\sqrt{\nu}\mathbf{B}(\mathbf{x}(t),t)d\mathbf{W}(t),\label{eq:Ito}
\end{equation}
where $\mathbf{x}(t)\in\mathbb{R}^{n}$ is the random position vector
of a particle at time $t$; $\mathbf{v}(\mathbf{x},t)$ denotes
the incompressible, deterministic drift in the particle motion; and
$\mathbf{W}(t)$ in an $m$-dimensional Wiener process with diffusion
matrix $\sqrt{\nu}\mathbf{B}(\mathbf{x},t)\in\mathbb{R}^{n\times m}$.
Here the dimensionless, nonsingular diffusion structure matrix $\mathbf{B}$
is $\mathcal{O}(1)$ with respect to the small parameter $\nu>0$. 

Let $p(\mathbf{x},t;\mathbf{x}_{0},t_{0})$ denote the probability
density function (PDF) for the current particle position $\mathbf{x}(t)$
with initial condition $\mathbf{x}_{0}(t_{0})=\mathbf{x}_{0}$. This
PDF is known to satisfy the classic Fokker-Planck equation \cite{risken84}
\begin{equation}
p_{t}+\mathbf{\mathbf{\bm{\nabla}}\cdot}\left(p\mathbf{v}\right)=\nu\tfrac{1}{2}\mathbf{\mathbf{\bm{\nabla}}}\cdot\left[\mathbf{\mathbf{\bm{\nabla}}}\cdot\left(\mathbf{B}\mathbf{B}^{\top}p\right)\right].\label{eq:FP equation}
\end{equation}
We can rewrite \eqref{eq:FP equation} as 
\begin{equation}
p_{t}+\mathbf{\mathbf{\bm{\nabla}}\cdot}\left(p\mathbf{\tilde{v}}\right)=\nu\mathbf{\bm{\nabla}}\cdot\left(\tfrac{1}{2}\mathbf{B}\mathbf{B}^{\top}\mathbf{\bm{\nabla}}p\right),\quad\tilde{\mathbf{v}}=\mathbf{v}-\tfrac{\nu}{2}\mathbf{\bm{\nabla}}\cdot\left(\mathbf{B}\mathbf{B}^{\top}\right),\label{eq:FP1}
\end{equation}
which is of advection-diffusion-form, \eqref{eq:adv-diff},
if $\tilde{\mathbf{v}}$ is incompressible, i.e., if
\begin{equation}
\mathbf{\bm{\nabla}}\cdot\left[\mathbf{\bm{\nabla}}\cdot\left(\mathbf{B}(\mathbf{x},t)\mathbf{B}^{\top}(\mathbf{x},t)\right)\right]\equiv0.\label{eq:hom diff tensor}
\end{equation}
Assuming \eqref{eq:hom diff tensor} (which holds, e.g., for
homogeneous diffusion), we define the \emph{probabilistic transport
tensor $\mathbf{\bar{P}}_{t_{0}}^{t_{1}}$ }as the time-average of
\[
\mathbf{P}_{t_{0}}^{t_{1}}\coloneqq\frac{1}{2}\left[\mathbf{\bm{\nabla}}_{0}\mathbf{F}_{t_{0}}^{t}\right]^{-1}\mathbf{\mathbf{B}}\left(\mathbf{F}_{t_{0}}^{t},t\right)\mathbf{\mathbf{B}}^{\top}\left(\mathbf{F}_{t_{0}}^{t},t\right)\left[\mathbf{\bm{\nabla}}_{0}\mathbf{F}_{t_{0}}^{t}\right]^{-\top}.
\]
We then conclude that all our results on diffusive scalar transport
in a deterministic velocity field carry over automatically to particle
transport in the stochastic velocity field \eqref{eq:Ito} with the
substitution $\mathbf{\bar{T}}_{t_{0}}^{t_{1}}=\mathbf{\bar{P}}_{t_{0}}^{t_{1}}$.
Namely, we have
\begin{theorem}
With the substitution $\mathbf{E}_{\mathcal{T}_0}(\mathbf{x}_{0})=\mathbf{\bar{P}}_{t_{0}}^{t_{1}}-\mathcal{T}_0\mathbf{I}$
and under assumption \eqref{eq:hom diff tensor}, uniform barriers
and enhancers to the transport of the probability-density $p(\mathbf{x},t_{1};\mathbf{x}_{0},t_{0})$
in the stochastic velocity field \eqref{eq:Ito} are null-surfaces
satisfying Theorem \ref{thm1}.
\end{theorem}
This result enables a purely deterministic computation of observed
surfaces of particle accumulation  and particle
clearance without a Monte--Carlo simulation
for \eqref{eq:Ito}.

\section{Numerical implementation and example}

For a two-dimensional velocity field $\mathbf{v}(\mathbf{x},t)$
and diffusion-structure tensor $\mathbf{D}(\mathbf{x},t)$, the main
algorithmic steps in locating diffusion barriers over a time interval
$[t_{0},t_{1}]$ are as follows (cf.~\emph{SI Appendix S7} for more
detail and a simple example):
\begin{description}
\item [{(A1)}] Define a Lagrangian grid $\mathcal{G}_{0}$ of initial conditions; generate trajectories $\mathbf{x}(t,t_{0},\mathbf{x}_{0})$ of the
velocity field $\mathbf{v}(\mathbf{x},t)$ with initial conditions
$\mathbf{x}_{0}\in\mathcal{G}_{0}$ at time $t_{0}$. 
\item [{(A2)}] For all times $t\in[t_{0},t_{1}],$ compute the deformation
gradient $\bm{\nabla}_{0}\mathbf{F}_{t_{0}}^{t}(\mathbf{x}_{0})=\bm{\nabla}_{0}\mathbf{x}(t,t_{0},\mathbf{x}_{0})$
over the grid $\mathcal{G}_{0}$ by finite differencing in $\mathbf{x}_{0}$
(cf. \cite{haller15}). Then, compute the 
tensor field $\mathbf{\bar{C}}_{\mathbf{D}}$
in \eqref{eq:C_D}. 
\item [{(A3)}] Compute the diffusion-barrier-strength diagnostic $\mathrm{DBS}(\mathbf{x}_{0})=\mathrm{trace}\,\mathbf{\bar{C}}_{\mathbf{D}}(\mathbf{x}_{0})$.
Its ridges and trenches highlight the most influential diffusion barriers (backward-time fronts and jet cores, respectively) at time $t_{0}$. 
\item [{(A4)}] Compute eigenvalues $\lambda_{1}(\mathbf{x}_{0})$, $\lambda_{2}(\mathbf{x}_{0})$
and corresponding 
eigenvectors $\bm{\xi}_{1}(\mathbf{x}_{0})$, $\bm{\xi}_{2}(\mathbf{x}_{0})$
of $\mathbf{\bar{C}}_{\mathbf{D}}(\mathbf{x}_{0})$. Compute closed diffusion barriers
as limit cycles of 
\eqref{eq:eta field}.
Outermost members of the limit-cycle families mark diffusion-based
material vortex boundaries at time $t_{0}$.  
\item [{(A5)}] To locate time-$t$ positions of material diffusion barriers,
advect them using the flow map $\mathbf{F}_{t_{0}}^{t}$. 
\end{description}
For probabilistic diffusion barriers in the stochastic velocity
field \eqref{eq:Ito}, apply steps \textbf{A1}-\textbf{A5} after setting
$\mathbf{D}=\frac{1}{2}\mathbf{B}\mathbf{B}^{\top}$.
 
Our main example will illustrate steps (A1)-(A5) in the identification
of boundaries for the largest mesoscale eddies in the Southern Ocean.
Known as Agulhas rings, theses eddies are believed to contribute significantly
to global circulation and climate via the warm and salty water they
ought to carry \cite{beal11}. Several studies have sought to estimate material
transport via these eddies by determining their boundaries from different
material coherence principles, which all tend to give different results
\cite{haller13,froyland15,hadjighasem16a,wang16,haller16} Here, for
the first time, we locate the boundaries of Agulhas rings based on
the very principle that makes them significant: their role as universal
barriers to the diffusion of relevant ocean water attributes they
transport.

\begin{figure*}
\centering
\includegraphics{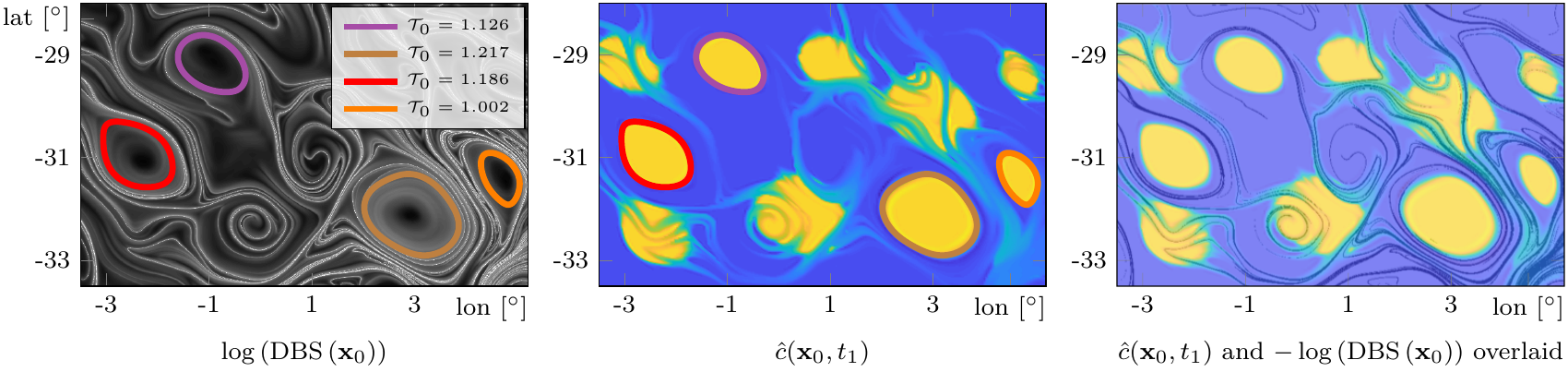}
\caption{Left: Predicted closed diffusion barriers overlaid on the
$\log(\mathrm{DBS}(\mathbf{x}_{0}))$ field; lighter colors mark higher DBS values. Middle: The diffused concentration,
$\hat{c}(\mathbf{x}_{0},t_{1})\coloneqq c(\mathbf{F}_{t_{0}}^{t_{1}}(\mathbf{x}_{0}),t_{1})$,
in Lagrangian coordinates $\mathbf{x}_{0}$; lighter colors mark higher concentration values; see also \emph{Supporting Animation SA1}. A The initial concentration
$c_{0}(\mathbf{x}_{0})$ is equal to one inside the predicted closed barriers and
inside seven shifted copies thereof, cf.~Fig.~\ref{fig:ocean-diffusion}, and to zero outside. 
Right: the ridges of $\log(\mathrm{DBS})$ overlaid on $\hat{c}(\mathbf{x}_{0},t_{1})$.}
\label{fig:ocean-combined} 
\end{figure*}

Figure \ref{fig:ocean-combined} shows diffusive coherent Agulhas ring
boundaries and surrounding diffusive barriers (backward-time fronts)
in the Southern Ocean, computed via steps (A1)-(A5) from satellite-altimetry-based 
surface velocities (cf.~\emph{SI Appendix S7} for more detail on the
data set). The predicted material ring boundaries are obtained as
described in step (A4). This prediction is confirmed 
by a diffusion simulation with Péclet number $Pe=\mathcal{O}(10^{4})$; see also
the Eulerian analogue in Fig.~S4 of the diffused 
concentration in \emph{Supporting Animation SA1}.
Figure \ref{fig:ocean-combined}c also confirms a similar barrier role for the ridges of
$\DBS(\mathbf{x}_{0})$ which closely align with observed
open barriers to diffusive transport. 

\begin{figure}
\centering
\includegraphics{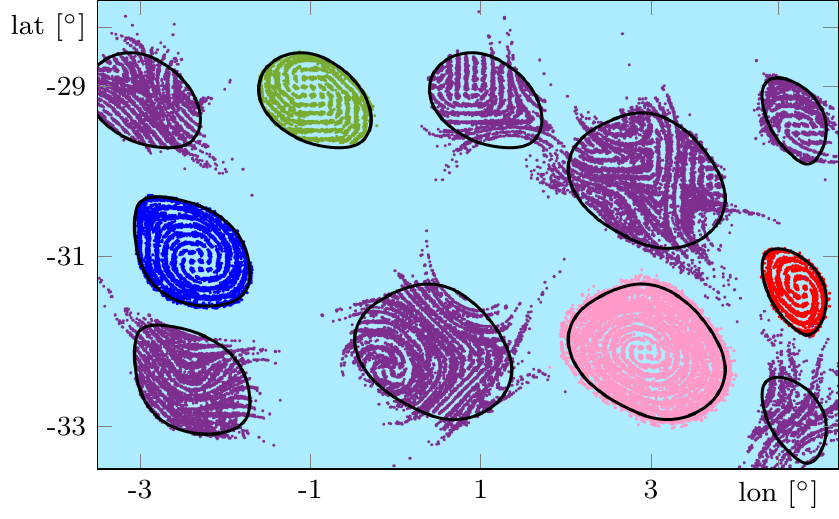}

\caption{Final positions of stochastic trajectories in the Lagrangian frame 
(cf. \eqref{eq:Lag_SDE}), initialized from the interiors of the closed black lines:
blue, green, pink and red are initialized within the closed diffusion barriers; purple ones are released from their translated copies for direct comparison. See \emph{Supporting Animation SA2} for the full animation in the Lagrangian frame and 
\emph{Supporting Animation SA3} in the physical (Eulerian) frame.}
\label{fig:ocean-diffusion} 
\end{figure}

Figure \ref{fig:ocean-diffusion} shows the final result of a Monte--Carlo simulation of
\eqref{eq:Ito} in the Lagrangian frame (cf.~\emph{SI Appendix S7}), given by
\[
d\mathbf{x}_0(t)=\sqrt{\nu}\mathbf{B}_0(\mathbf{x}_0(t),t)d\mathbf{W}(t),\mathbf{B}_0\coloneqq\left[\mathbf{\bm{\nabla}}_{0}\mathbf{F}_{t_{0}}^{t}\right]^{-1}\mathbf{B}\left(\mathbf{F}_{t_{0}}^{t},t\right),\label{eq:Lag_SDE}
\]
with homogeneous diffusion-structure matrix $\mathbf{B}=\mathbf{I}$,
whose Fokker--Planck equation coincides with the advection--diffusion equation
in our previous simulation.
The figure confirms the role of the ring boundaries (computed from the deterministic velocity field)
as sharp barriers to particle transport under uncertainties in the velocity field. We show the evolving Monte--Carlo simulation
in \emph{Supporting Animations SA2-SA3}.

\section{Conclusions}

We have pointed out that the presence of the slightest diffusion in a deterministic
flow yields an unambiguous, first-principles-based physical definition for transport
barriers as material surfaces that block diffusive transport the most efficiently. We
have found that in any dimension, such barriers lie close to minimizers of a
universal, non-dimensionalized transport functional that measures the leading-order
diffusive transport through material surfaces. 
Of these minimizers, a special set of most observable barriers is formed by those that maintain uniformly high concentration gradients, and hence uniform transport density, along themselves.  
Even such uniform barriers, however, will generally differ from coherent structures identified from purely advective considerations (Remark 1). Beyond the exact differential equations describing transport barriers, we have obtained a predictive diagnostic field, $\mathrm{DBS} (\mathbf{x}_0)$, that signals barrier location and strength from purely advective computations (Remark 2). Finally, we have discussed how the proposed methodology identifies probabilistic material barriers and enhancers to particle transport in multi-dimensional stochastic velocity fields.

Our results identify the main enhancers and inhibitors of transport
in diffusive and random flows without costly numerical solutions of
PDEs or Monte-Carlo simulations of stochastic flow models. By construction,
the structures we obtain are robust with respect to small diffusive
effects, including measurement uncertainties in observational velocity
data or modeling errors in numerically generated velocity fields.
Our detection scheme for transport extremizers is independent of the local availability of the diffusive tracer and of the initial distribution of its gradient field. The theoretically optimal transport extremizers identified
here should also be useful as benchmarks for the development for future
diagnostics targeting transport barriers in sparse data. Further theoretical work is required for a more detailed classification
of diffusion extremizers in higher dimensions and in compressible flows. On the computational
side, the accurate identification of diffusion extremizers identified
here requires efficient numerical schemes for null-surfaces. On the
applications side, further examples of practically relevant and multi-scale
velocity fields need to be analyzed in detail to assess further practical implications
of the barrier-detection method introduced here. 

\subsubsection*{Acknowledgements}

We are grateful to R Abernathey, FJ Beron--Vera, T Breunung, S Katsanoulis, A Constantin, M Mathur, G Pavliotis,  M Rubin and J-L Thiffeault for useful discussions and comments, and to N Schilling for contributions to the animation code. GH and DK acknowledge support from the Turbulent Superstructures priority program of the German National Science Foundation (DFG).

\showacknow{} 

\vfil\eject
\subsection*{S1: Expansion of the total transport in $\nu$}

We denote the restriction of the concentration field $c(\mathbf{x},t)$
to trajectories of the velocity field by $\hat{c}(\mathbf{x}_{0},t)=c\left(\mathbf{F}_{t_{0}}^{t}(\mathbf{x}_{0}),t\right)$. We then use the advection-diffusion equation to conclude that the
time-derivative of $\hat{c}(\mathbf{x}_{0},t)$ satisfies 
\begin{equation}
\partial_{t}\hat{c}(\mathbf{x}_{0},t)=\nu\mathbf{\mathbf{\bm{\nabla}}}\cdot\left(\mathbf{D}\left(\mathbf{F}_{t_{0}}^{t}(\mathbf{x}_{0}),t\right)\mathbf{\bm{\nabla}}\left(c\left(\mathbf{F}_{t_{0}}^{t}(\mathbf{x}_{0}),t\right)\right)\right).\label{eq:partial c hat classic 1}
\end{equation}
Introducing the Lagrangian diffusion structure tensor $\hat{\mathbf{D}}\left(\mathbf{x}_{0},t\right)=\mathbf{D}\left(\mathbf{F}_{t_{0}}^{t}(\mathbf{x}_{0}),t\right)$,
we can rewrite \eqref{eq:partial c hat classic 1} as 
\begin{equation}
\partial_{t}\hat{c}=\nu\mathbf{\mathbf{\bm{\nabla}}}\cdot\left(\hat{\mathbf{D}}\mathbf{\mathbf{\bm{\nabla}}}\hat{c}\right).\label{eq:partial c hat classic 2}
\end{equation}
A lengthy calculation leads to the Lagrangian form of the advection-diffusion
equation as \cite{press81,knobloch92,tang96,thiffeault03} 
\begin{equation}
\partial_{t}\hat{c}=\nu\mathbf{\mathbf{\bm{\nabla}}}_{0}\cdot\left(\left[\mathbf{\mathbf{\bm{\nabla}}}\mathbf{F}_{t_{0}}^{t}\right]^{-1}\hat{\mathbf{D}}\left[\mathbf{\nabla}\mathbf{F}_{t_{0}}^{t}\right]^{-\top}\mathbf{\mathbf{\bm{\nabla}}}_{0}\hat{c}\right).\label{eq:comp Lagr correct-1}
\end{equation}
Taking Lagrangian spatial gradient $\mathbf{\bm{\nabla}}_{0}$ of
both sides and integrating in time, we obtain 
\begin{equation}
\mathbf{\mathbf{\bm{\nabla}}}_{0}\hat{c}=\mathbf{\mathbf{\bm{\nabla}}}_{0}c_{0}+\nu\int_{t_{0}}^{t_{1}}\mathbf{\mathbf{\bm{\nabla}}}_{0}\left[\mathbf{\mathbf{\bm{\nabla}}}_{0}\cdot\left(\mathbf{T}_{t_{0}}^{t}\mathbf{\mathbf{\bm{\nabla}}}_{0}\hat{c}\right)\right]dt.\label{eq:grad c hat}
\end{equation}
Substitution of \eqref{eq:grad c hat} into the definition of $\Sigma_{t_{0}}^{t_{1}}$
then gives 
\begin{align*}
 & \Sigma_{t_{0}}^{t_{1}}(\mathcal{M}_{0})=\nu\int_{t_{0}}^{t_{1}}\int_{\mathcal{M}_{0}}\left[\mathbf{\mathbf{\bm{\nabla}}}_{0}c_{0}\right]^{\top}\mathbf{T}_{t_{0}}^{t}\mathbf{n}_{0}dA_{0}\,dt\\
 & +\nu^{2}\int_{t_{0}}^{t_{1}}\int_{\mathcal{M}_{0}}\left[\int_{t_{0}}^{t}\mathbf{\mathbf{\bm{\nabla}}}_{0}\left[\mathbf{\mathbf{\bm{\nabla}}}_{0}\cdot\left(\mathbf{T}_{t_{0}}^{s}\mathbf{\mathbf{\bm{\nabla}}}_{0}\hat{c}\right)\right]ds\right]^{\top}\mathbf{T}_{t_{0}}^{t}\mathbf{n}_{0}dA_{0}\,dt.
\end{align*}
We will now prove that the second term in this equation is of order
$o(\nu)$, i.e.,

{\small{}{} 
\begin{equation}
\lim_{\nu\to0}\nu\int_{t_{0}}^{t_{1}}\int_{\mathcal{M}_{0}}\left[\int_{t_{0}}^{t}\mathbf{\mathbf{\bm{\nabla}}}_{0}\left[\mathbf{\mathbf{\bm{\nabla}}}_{0}\cdot\left(\mathbf{T}_{t_{0}}^{s}\mathbf{\mathbf{\bm{\nabla}}}_{0}\hat{c}\right)\right]ds\right]^{\top}\mathbf{T}_{t_{0}}^{t}\mathbf{n}_{0}dA_{0}\,dt=0.\label{eq:ansatz}
\end{equation}
}{\small \par}

To this end, we need estimates on the solution of \eqref{eq:comp Lagr correct-1},
which we rewrite here using the tensor $\mathbf{T}_{t_{0}}^{t}$ as
\begin{align}
\partial_{t}\hat{c}(\mathbf{x}_{0},t) & =\nu\mathbf{\mathbf{\bm{\nabla}}}_{0}\cdot\left(\mathbf{T}_{t_{0}}^{t}(\mathbf{x}_{0})\mathbf{\mathbf{\bm{\nabla}}}_{0}\hat{c}(\mathbf{x}_{0},t)\right),\quad(\mathbf{x}_{0},t)\in U\times\left[t_{0},t_{1}\right],\label{main}\\
\hat{c}(\mathbf{x}_{0},t_{0}) & =\hat{c}(\mathbf{x}_{0}),\nonumber 
\end{align}
By our assumption of Hölder continuity for $\mathbf{D}$ and smoothness
for all other quantities involved, we obtain that $\mathbf{T}_{t_{0}}^{t}(\mathbf{x}_{0})$
is Hölder continuous. Specifically, for any entry $T_{ij}(\mathbf{x}_{0},t):=\left[\mathbf{T}_{t_{0}}^{t}(\mathbf{x}_{0})\right]_{ij}$
of the matrix representation of $\mathbf{T}_{t_{0}}^{t}$, we have
the bounds 
\begin{equation}
\begin{split} & \left|T_{ij}(\mathbf{x}_{0},t)-T_{ij}(\mathbf{y}_{0},s)\right|\leq C_{1}\left|\mathbf{x}_{0}-\mathbf{y}_{0}\right|^{\alpha}+C_{2}\left|t-s\right|^{\frac{\alpha}{2}},\\
 & \left|\bm{\nabla}_{0}T_{ij}(\mathbf{x}_{0},t)-\bm{\nabla}_{0}T_{ij}(\mathbf{y}_{0},t)\right|\leq C_{3}|\mathbf{x}_{0}-\mathbf{y}_{0}|^{\alpha},
\end{split}
\label{Hold}
\end{equation}
for some constant $0<\alpha\leq1$ and for all $\mathbf{x}_{0},\mathbf{y}_{0}\in U$
and $t,s\in[t_{0},t_{1}]$. By the positive definiteness of $\mathbf{T}_{t_{0}}^{t}(\mathbf{x}_{0})$,
we also have 
\begin{equation}
\lambda|\mathbf{u}|^{2}\leq\left\langle \mathbf{u},\mathbf{T}_{t_{0}}^{t}(\mathbf{x}_{0})\mathbf{u}\right\rangle \leq\Lambda|\mathbf{u}|^{2},\quad\mathbf{u}\in\mathbb{R}^{n},\,\,\mathbf{x}_{0}\in U,\,\,t\in[t_{1},t_{2}],\label{ell}
\end{equation}
which implies the bounds 
\begin{equation}
\frac{|\mathbf{u}|^{2}}{\Lambda}\leq\left\langle \mathbf{u},\left[\mathbf{T}_{t_{0}}^{t}(\mathbf{x}_{0})\right]^{-1}\mathbf{u}\right\rangle \leq\frac{|\mathbf{u}|^{2}}{\lambda},\quad\lambda^{n}\leq\det\mathbf{T}_{t_{0}}^{t}(\mathbf{x}_{0})\leq\Lambda^{n},\label{Ainv}
\end{equation}
for all $\mathbf{u}\in\mathbb{R}^{n},\,\,\mathbf{x}_{0}\in U$ and
$t\in[t_{1},t_{2}]$. Next, we observe that \eqref{eq:ansatz} is
satisfied when 
\begin{equation}
\sup_{\mathbf{x}_{0}\in U,t\in[t_{0},t_{1}]}\left|\mathbf{\mathbf{\bm{\nabla}}}_{0}\hat{c}(\mathbf{x}_{0},t)-\mathbf{\mathbf{\bm{\nabla}}}_{0}c_{0}(\mathbf{x}_{0})\right|=\mathcal{O}(\nu^{q}),\label{sup}
\end{equation}
holds for some $q>0$, as one obtains using \eqref{eq:grad c hat}
and estimating the supremum norm in $\mathbf{x}_{0}$ and $t$ using
\eqref{Hold}. Using the assumption that $c_{0}\in C^{2}(U),$ we
will now show that \eqref{sup} holds, and hence \eqref{eq:ansatz}
is indeed satisfied. In our presentation, we will utilize a scaling
approach described in \cite{friedman13}.

Introducing the rescaled time variable $\tau:=\nu(t-t_{0})$ as well
as the shifted and rescaled concentration $w(\mathbf{x}_{0},\tau):=\hat{c}(\mathbf{x}_{0},t_{0}+\frac{\tau}{\nu})-c_{0}(\mathbf{x}_{0})$,
then setting $\mathbf{T}_{\nu}(\mathbf{x}_{0},\tau):=\mathbf{T}_{t_{0}}^{t_{0}+\frac{\tau}{\nu}}(\mathbf{x}_{0})$,
we can rewrite \eqref{main} as 
\begin{equation}
\begin{cases}
w_{\tau}=\mathbf{T}_{\nu}\cdot\mathbf{\mathbf{\bm{\nabla}}}_{0}^{2}w+\left(\mathbf{\mathbf{\bm{\nabla}}}_{0}\cdot\mathbf{T}_{\nu}\right)\bm{\nabla}_{0}w+\mathbf{\mathbf{\bm{\nabla}}}_{0}\cdot\left(\mathbf{T}_{\nu}\bm{\nabla}_{0}c_{0}\right),\\
w(\mathbf{x}_{0},0)=0,\quad(\mathbf{x}_{0},\tau)\in U\times[0,\nu(t_{1}-t_{0})].
\end{cases}\label{mainw}
\end{equation}
Condition \eqref{sup} is then equivalent to 
\begin{equation}
\sup_{\mathbf{x}_{0}\in\Omega,t\in[0,\tau_{1}]}|\mathbf{\mathbf{\bm{\nabla}}}_{0}w(\mathbf{x}_{0},\tau)|=\mathcal{O}(\nu^{q}),\qquad\tau_{1}:=\nu(t_{1}-t_{0})\label{supw}
\end{equation}
for some $q>0$. Let 
\begin{align}
Z(\mathbf{x}_{0},\tau;\bm{\xi},s): & =\frac{\exp\left[-\frac{\left\langle \mathbf{x}_{0}-\bm{\xi},\mathbf{T}_{\nu}^{-1}(\bm{\xi},s)(\mathbf{x}_{0}-\bm{\xi})\right\rangle }{4(\tau-s)}\right]}{(2\sqrt{\pi})^{n}\left[\det\mathbf{T}_{\nu}(\bm{\xi},s)\right]{}^{\frac{1}{2}}(\tau-s)^{\frac{n}{2}}},\label{defZ}\\
Z_{\tau} & =\mathbf{T}_{\nu}\cdot\mathbf{\mathbf{\bm{\nabla}}}_{0}^{2}Z,\nonumber 
\end{align}
for $\mathbf{x}_{0},\bm{\xi}\in\Omega$ and $\tau,s\in[0,\tau_{1}]$,
denote the fundamental solution of the homogeneous, second-order part
of \eqref{mainw}. For later computations, we note that with the $n$-dimensional
volume element $d\bm{\xi}=d\xi_{1}...d\xi_{n}$, we have the estimate{\tiny{}{}
\begin{align}
 & \int_{\Omega}Z(\mathbf{x}_{0},\tau;\mathbf{\bm{\xi}},s)\,d\bm{\xi}\nonumber \\
 & =\int_{\Omega}(2\sqrt{\pi})^{-n}\left[\det\mathbf{T}_{\nu}^{-1}\right]^{-\frac{1}{2}}(\tau-s)^{-\frac{n}{2}}e^{-\frac{\left\langle \mathbf{x}_{0}-\bm{\xi},\mathbf{T}_{\nu}^{-1}(\mathbf{x}_{0}-\bm{\xi})\right\rangle }{4(\tau-s)}}\,d\bm{\xi}\nonumber \\
 & \leq\int_{\Omega}(2\sqrt{\pi})^{-n}\lambda^{-\frac{n}{2}}(\tau-s)^{-\frac{n}{2}}e^{-\frac{|\mathbf{x}_{0}-\bm{\mathbf{\xi}}|^{2}}{4\Lambda(\tau-s)}}\,d\mathbf{\bm{\xi}},\label{estZ}
\end{align}
}where we have used the inequalities in \eqref{Ainv}. With the rescaled
spatial variable $\mathbf{y}$ and the rescaled volume form $d\mathbf{y}$
defined as 
\begin{equation}
\begin{split} & \mathbf{y}=(2\Lambda)^{-\frac{1}{2}}(\tau-s)^{-1/2}(\mathbf{x}-\bm{\xi}),\quad d\mathbf{y}=(2\Lambda)^{-\frac{n}{2}}(\tau-s)^{-\frac{n}{2}}d\mathbf{\bm{\xi}},\\
\\
\end{split}
\label{scaling}
\end{equation}
we define the set $\Omega_{\mathbf{x}_{0},\tau,s}:=(2\Lambda)^{-\frac{1}{2}}(\tau-s)^{-1/2}(\mathbf{x}_{0}-\Omega)$
to obtain from \eqref{estZ} the estimate 
\begin{equation}
\begin{split}\int_{\Omega}Z(\mathbf{x}_{0},\tau;\mathbf{\bm{\xi}},s)\,d\bm{\xi} & \leq\pi^{-\frac{n}{2}}\left(\frac{\Lambda}{\lambda}\right)^{\frac{n}{2}}\int_{\Omega_{\mathbf{x},\tau,s}}e^{-|\mathbf{y}|^{2}}d\mathbf{y}\\
 & \leq\pi^{-\frac{n}{2}}\left(\frac{\Lambda}{\lambda}\right)^{\frac{n}{2}}\int_{\mathbb{R}^{n}}e^{-|\mathbf{y}|^{2}}d\mathbf{y}=\left(\frac{\Lambda}{\lambda}\right)^{\frac{n}{2}},
\end{split}
\label{estZ2}
\end{equation}
where we have used that $\int_{-\infty}^{\infty}e^{-r^{2}}\,dr=\sqrt{\pi}$.
We also recall from \cite{friedman13} (Theorem 3, p. 8), that for
any continuous function $f:\Omega\times[0,\tau_{1}]\to\mathbb{R}$,
the integral 
\begin{equation}
V(\mathbf{x}_{0},\tau):=\int_{0}^{\tau}\int_{\Omega}Z(\mathbf{x}_{0},\tau;\mathbf{\bm{\xi}},s)f(\bm{\xi},s)\,d\bm{\xi}ds\label{volpot}
\end{equation}
is continuously-differentiable with respect to $\mathbf{x}_{0}$ and
satisfies 
\begin{equation}
\mathbf{\mathbf{\bm{\nabla}}}_{0}V(\mathbf{x}_{0},\tau)=\int_{0}^{\tau}\int_{\Omega}\mathbf{\mathbf{\bm{\nabla}}}_{0}Z(\mathbf{x}_{0},\tau;\bm{\xi},s)f(\bm{\xi},s)\,d\bm{\xi}ds.\label{derpot}
\end{equation}
\\
 As shown in \cite{friedman13} (Theorem 9, p.21), the variation of
constants formula applied to \eqref{mainw} gives its solution in
the form {\footnotesize{}{} 
\begin{equation}
\begin{split} & w(\mathbf{x}_{0},\tau)=\int_{0}^{\tau}\int_{\Omega}Z\mathbf{\mathbf{\bm{\nabla}}}_{0}\cdot\left(\mathbf{T}_{\nu}\bm{\nabla}_{0}c_{0}\right)\,d\bm{\xi}\,ds\\
 & +\int_{0}^{\tau}\int_{\Omega}Z(\mathbf{x}_{0},\tau;\mathbf{\bm{\xi}},s)\times\\
 & \,\,\,\,\,\,\,\times\left(\int_{0}^{s}\int_{\Omega}\Phi\left(\bm{\xi},s;\bm{\eta},\sigma\right)\left(\mathbf{T}_{\nu}(\bm{\eta},\sigma)\bm{\nabla}_{0}c_{0}(\bm{\eta})\right)\,d\bm{\eta}\,d\sigma\right)\,d\bm{\xi}\,ds\\
 & =:W_{1}(\mathbf{x}_{0},\tau)+W_{2}(\mathbf{x}_{0},\tau),
\end{split}
\label{Fund}
\end{equation}
}for some (not explicitly known) function $\Phi$ that satisfies the
estimate 
\begin{equation}
\left|\Phi\left(\bm{\xi},s;\bm{\eta},\sigma\right)\right|\leq C_{4}\frac{1}{|s-\sigma|^{\mu}}\frac{1}{|\xi-\eta|^{n+2-2\mu-\alpha}},\label{phi}
\end{equation}
for any constant $\mu\in\left(1-\frac{\alpha}{2},1\right)$, where
$\alpha$ is the Hölder-exponent in \eqref{Hold}.

\noindent To estimate the spatial gradient of $W_{1}$, we use the
formula for the $\mathbf{x}_{0}$-derivative of \eqref{Fund} in \eqref{derpot}
to obtain{\footnotesize{}{} 
\begin{equation}
\begin{split} & \left|\bm{\nabla}_{0}W_{1}\right|=\left|\bm{\nabla}_{0}\int_{0}^{\tau}\int_{\Omega}Z\mathbf{\mathbf{\bm{\nabla}}}_{0}\cdot\left(\mathbf{T}_{\nu}\bm{\nabla}_{0}c_{0}\right)\,d\bm{\xi}\,ds\right|\\
 & =\left|\int_{0}^{\tau}\int_{\Omega}\left(\bm{\nabla}_{0}Z\right)\mathbf{\mathbf{\bm{\nabla}}}_{0}\cdot\left(\mathbf{T}_{\nu}\bm{\nabla}_{0}c_{0}\right)\,d\bm{\xi}\,ds\right|\\
 & \leq\int_{0}^{\tau}\int_{\Omega}\frac{1}{2|\tau-s|}\left|\mathbf{T}_{\nu}^{-1}(\bm{\xi},s)(\mathbf{x}_{0}-\bm{\xi})\right|\left|Z\right|\left|\mathbf{\mathbf{\bm{\nabla}}}_{0}\cdot\left(\mathbf{T}_{\nu}\bm{\nabla}_{0}c_{0}\right)\right|\,d\bm{\xi}\,ds,
\end{split}
\label{essW1}
\end{equation}
}where we also used the definition \eqref{defZ} in evaluating $\bm{\nabla}_{0}Z$.
From \eqref{Ainv}, we obtain $\|\mathbf{T}_{\nu}^{-1}\|=\lambda^{-1}$,
and hence we can further write \eqref{essW1} as {\footnotesize{}{}
\begin{equation}
\begin{split}\left|\bm{\nabla}_{0}W_{1}\right| & \leq\frac{1}{\lambda}\int_{0}^{\tau}\int_{\Omega}\frac{\left|Z\right|}{2\left|\tau-s\right|}\left|\mathbf{\mathbf{\bm{\nabla}}}_{0}\cdot\left(\mathbf{T}_{\nu}\bm{\nabla}_{0}c_{0}\right)\right|\,d\bm{\xi}\,ds\\
 & \leq\frac{\|\mathbf{\mathbf{\bm{\nabla}}}_{0}\cdot\left(\mathbf{T}_{\nu}\bm{\nabla}_{0}c_{0}\right)\|_{C^{0}(\Omega)}}{\lambda}\int_{0}^{\tau}\int_{\Omega}\frac{1}{2|\tau-s|}\left|\mathbf{x}_{0}-\bm{\xi}\right|\left|Z\right|\,d\bm{\xi}\,ds\\
 & \leq C_{5}\frac{\|c_{0}\|_{C^{2}(\Omega)}}{\lambda}\int_{0}^{\tau}\int_{\Omega}\frac{1}{2|\tau-s|}\left|\mathbf{x}_{0}-\bm{\xi}\right|\left|Z\right|\,d\bm{\xi}\,ds.
\end{split}
\label{eq:w1est1}
\end{equation}
}Next, as in the calculation of the integral in \eqref{estZ}, we
use the scaling \eqref{scaling} in \eqref{eq:w1est1} to obtain 
\begin{equation}
\begin{split}\left|\bm{\nabla}_{0}W_{1}\right| & \leq C_{5}\frac{\Lambda\|c_{0}\|_{C^{2}(\Omega)}}{\lambda}\int_{0}^{\tau}\frac{1}{\sqrt{\tau-s}}\left(\int_{\mathbb{R}^{n}}|\bm{y}|e^{-|\bm{y}|^{2}}\,d\bm{y}\right)\,ds\\
 & \leq C_{6}\frac{\Lambda\|c_{0}\|_{C^{2}(\Omega)}}{\lambda}\int_{0}^{\tau}\frac{1}{\sqrt{\tau-s}}\,ds\\
 & \leq C_{7}\sqrt{\tau}=\mathcal{O}\left(\nu^{\frac{1}{2}}\right).
\end{split}
\label{eq:w1estfinal}
\end{equation}
To estimate the spatial gradient of $W_{2}$ in \eqref{Fund}, we
proceed similarly by using the growth condition \eqref{phi} to obtain
{\tiny{}{} 
\begin{equation}
\begin{split} & \left|\bm{\nabla}_{0}W_{2}\right|\\
 & \leq\int_{0}^{\tau}\int_{\Omega}\frac{1}{2\left|\tau-s\right|}\left|\mathbf{T}_{\nu}^{-1}(\mathbf{x}_{0}-\bm{\xi})\right|\left|Z\right|\times\\
 & \,\,\,\,\,\,\,\,\,\,\,\,\,\,\,\left(\int_{0}^{s}\int_{\Omega}\left|\Phi\right|\left|\mathbf{\mathbf{\bm{\nabla}}}_{0}\cdot\left(\mathbf{T}_{\nu}\bm{\nabla}_{0}c_{0}\right)\right|\,d\bm{\eta}\,d\sigma\right)\,d\bm{\xi}\,ds\\
 & \leq C_{8}\frac{\|\mathbf{\mathbf{\bm{\nabla}}}_{0}\cdot\left(\mathbf{T}_{\nu}\bm{\nabla}_{0}c_{0}\right)\|_{C^{0}(\Omega)}}{\lambda}\times\\
 & \,\,\,\,\,\times\int_{0}^{\tau}\int_{\Omega}\frac{1}{2|\tau-s|}\left|\mathbf{x}_{0}-\bm{\xi}\right|\left|Z\right|\times\\
 & \,\,\,\,\,\,\,\,\,\,\,\,\,\,\,\left(\int_{0}^{s}\frac{d\sigma}{|s-\sigma|^{\mu}}\int_{\Omega}\frac{d\bm{\eta}}{|\bm{\xi}-\bm{\eta}|^{n+2-2\mu-\alpha}}\right)\,d\bm{\xi}\,ds.
\end{split}
\label{eq:w2est1}
\end{equation}
}Since $\Omega$ is bounded, there exists a ball of radius $R$ such
that $\Omega+\Omega\subset B_{R}$ and therefore, noticing that $2-2\mu-\alpha>0$
by $1-\frac{\alpha}{2}<\mu<1$, we find that 
\begin{equation}
\int_{\Omega}\frac{d\bm{\eta}}{|\bm{\xi}-\bm{\eta}|^{n+2-2\mu-\alpha}}\leq C_{9}\left.r^{2-2\mu-\alpha}\right|_{r=0}^{r=R}=C_{9}R^{2-2\mu-\alpha}.
\end{equation}
As in \eqref{essW1}, we can estimate the integral of $\left|\mathbf{x}_{0}-\bm{\xi}\right|\left|Z\right|$
to obtain {\scriptsize{}{} 
\begin{equation}
\begin{split} & \left|\bm{\nabla}_{0}W_{2}\right|\\
 & \leq C_{9}\frac{R^{2-2\mu-\alpha}\|u_{0}\|_{C^{2}(\Omega)}}{\lambda}\times\\
 & \,\,\,\,\,\,\,\times\int_{0}^{\tau}\int_{\Omega}\frac{c}{2|\tau-s|}\left|\mathbf{x}_{0}-\bm{\xi}\right|\left|Z\right|\left(\int_{0}^{s}\frac{d\sigma}{|s-\sigma|^{\mu}}\right)\,d\bm{\xi}\,ds\\
 & \leq C_{10}\frac{R^{2-2\mu-\alpha}\|c_{0}\|_{C^{2}(\Omega)}}{\lambda}\int_{0}^{\tau}\frac{1}{\sqrt{\tau-s}}\left(\int_{0}^{s}\frac{1}{|s-\sigma|^{\mu}}\,d\sigma\right)\,ds\\
 & \leq C_{11}\int_{0}^{\tau}\frac{|\tau-s|^{1-\mu}}{\sqrt{\tau-s}}\,ds\\
 & \leq C_{12}|\tau|^{\frac{3}{2}-\mu}=\mathcal{O}\left(\nu^{\frac{\alpha+1}{2}}\right).
\end{split}
\label{eq:w2estfinal}
\end{equation}
}The estimates \eqref{eq:w1estfinal}-\eqref{eq:w2estfinal} together
prove \eqref{supw}, which then implies \eqref{sup}, which in turn
implies \eqref{eq:ansatz}, as claimed.

\subsection*{S2: Objectivity of the transport tensor\label{appendix: objectivity}}

Physically, the Eulerian flux density $\Phi\left(\mathbf{x},t\right)=\nu\mathbf{D}\left(\mathbf{x},t\right)\bm{\nabla_{x}}c\left(\mathbf{x},t\right)\cdot\mathbf{n}dA$
at a point $\mathbf{x}$ at time $t$ through a surface element $dA$
with unit normal $\mathbf{n}\left(\mathbf{x},t\right)$ must be independent
of rotations and translations of observer. Consequently, under an
observer change 
\begin{equation}
\mathbf{x}=\mathbf{Q}(t)\mathbf{y}+\mathbf{b}(t),\qquad\mathbf{Q}(t_{0})=\mathbf{I},\label{eq:observer change}
\end{equation}
we must have $\Phi\left(\mathbf{x},t\right)=\Phi\left(\mathbf{Q}(t)\mathbf{y}+\mathbf{b}(t),t\right),$
and hence 
\begin{align*}
 & \nu\mathbf{D}\left(\mathbf{x},t\right)\bm{\nabla_{x}}c\left(\mathbf{x},t\right)\cdot\mathbf{n}\left(\mathbf{x},t\right)dA\\
 & =\nu\mathbf{D}\left(\mathbf{Q}(t)\mathbf{y}+\mathbf{b}(t),t\right)\mathbf{Q}(t)\bm{\nabla_{y}}c\left(\mathbf{Q}(t)\mathbf{y}+\mathbf{b}(t),t\right)\cdot\mathbf{Q}(t)\tilde{\mathbf{n}}\left(\mathbf{y},t\right)dA\\
 & =\nu\tilde{\mathbf{D}}\left(\mathbf{y},t\right)\bm{\nabla_{y}}\tilde{c}\left(\mathbf{y},t\right)\cdot\tilde{\mathbf{n}}\left(\mathbf{y},t\right)dA,
\end{align*}
where we have defined the transformed diffusion tensor 
\begin{equation}
\tilde{\mathbf{D}}\left(\mathbf{y},t\right)=\mathbf{Q}^{\top}(t)\mathbf{D}\left(\mathbf{x},t\right)\mathbf{Q}(t),\label{eq:D transformation  formula}
\end{equation}
and used the fact that the area element $dA$ remains unchanged under
rigid-body rotations and translations embodied by \eqref{eq:observer change}.
Using \eqref{eq:D transformation  formula} together with $\mathbf{\bm{\nabla}}_{0}\mathbf{F}_{t_{0}}^{t}\left(\mathbf{x}_{0}\right)=\mathbf{Q}(t)\mathbf{\bm{\nabla}}_{0}\mathbf{\tilde{F}}_{t_{0}}^{t}\left(\mathbf{y}_{0}\right)$
in the definition of $\mathbf{T}_{t_{0}}^{t}$ gives 
\begin{align*}
\mathbf{T}_{t_{0}}^{t} & \left(\mathbf{x}_{0}\right)=\left[\mathbf{\bm{\nabla}}_{0}\mathbf{F}_{t_{0}}^{t}\left(\mathbf{x}_{0}\right)\right]^{-1}\mathbf{D}\left(\mathbf{F}_{t_{0}}^{t}\left(\mathbf{x}_{0}\right),t\right)\left[\mathbf{\bm{\nabla}}_{0}\mathbf{F}_{t_{0}}^{t}\left(\mathbf{x}_{0}\right)\right]^{-\top}\\
 & =\left[\mathbf{\bm{\nabla}}_{0}\mathbf{\tilde{F}}_{t_{0}}^{t}\left(\mathbf{y}_{0}\right)\right]^{-1}\mathbf{\mathbf{\tilde{D}}}\left(\mathbf{\tilde{F}}_{t_{0}}^{t},t\right)\left[\mathbf{\bm{\nabla}}_{0}\mathbf{\tilde{F}}_{t_{0}}^{t}\left(\mathbf{y}_{0}\right)\right]^{-\top}=\mathbf{\tilde{T}}_{t_{0}}^{t}\left(\mathbf{y}_{0}\right).
\end{align*}
This then proves the frame-indifference of the transport tensor $\mathbf{\bar{T}}_{t_{0}}^{t_{1}}\left(\mathbf{x}_{0}\right)$.
as a tensor acting on, and mapping back to, the initial configuration,
which is unaffected by the frame change.

\subsection*{S3: Reformulation of the transport functional}

Under a general parametrization $\mathbf{x}_{0}(\mathbf{s})$ of $\mathcal{M}_{0}$,
the integral in the functional $\mathcal{E}_{\mathcal{T}}$ can be
rewritten as 
\begin{equation}
\int_{\mathcal{M}_{0}}\left[\left\langle \mathbf{n}_{0},\mathbf{\bar{T}}_{t_{0}}^{t_{1}}\mathbf{n}_{0}\right\rangle -\mathcal{T}\right]\sqrt{\det\mathbb{{\bf G}}}\,ds_{1}\ldots ds_{n-1},\label{eq: transport S21}
\end{equation}
where $G_{ij}\left(\partial_{\mathbf{s}}\mathbf{x}_{0}(\mathbf{s})\right)=\left\langle \frac{\partial\mathbf{x}_{0}}{\partial s_{i}},\frac{\partial\mathbf{x}_{0}}{\partial s_{j}}\right\rangle $,
$i,j=1,\ldots,n-1$, denotes the $\left(i,j\right)$ entry of the
Gramian matrix $\mathbb{{\bf G}}\left(\partial_{\mathbf{s}}\mathbf{x}_{0}(\mathbf{s})\right)$
of the parametrization, with $\sqrt{\det\mathbb{{\bf G}}\left(\partial_{\mathbf{s}}\mathbf{x}_{0}(\mathbf{s})\right)}ds_{1}\ldots ds_{n-1}$
providing the surface are element on $\mathcal{M}_{0}$.

To express the integrand of \eqref{eq: transport S21} fully in terms
of tangent vectors $\partial_{s_{i}}\mathbf{x}_{0}(\mathbf{s})$,
we first consider a general invertible linear operator $\mathbf{A}:\mathbb{R}^{n\times n}\to\mathbb{R}^{n\times n},$
and a unit vector $\mathbf{n}_{0}$ selected to be normal to an $(n-1)$
dimensional hyperplane $E=\mathrm{span}\left\{ \mathbf{u}_{1},\ldots,\mathbf{u}_{n-1}\right\} $
of $n-1$ linearly independent vectors $\mathbf{u}_{i}\in\mathbb{R}^{n}$.
Recall that the $n-1$-dimensional area of the parallelepiped spanned
by these vectors is equal to 
\[
\mathrm{area}(\mathbf{u}_{1},\ldots,\mathbf{u}_{n-1})=\sqrt{\det\mathbf{G}(\mathbf{u}_{1,}\ldots,\mathbf{u}_{n-1})},
\]
with the entries of the Gramian matrix $\mathbf{G}$ defined as $G_{ij}=\left\langle \mathbf{u}_{i},\mathbf{u}_{j}\right\rangle .$
Similarly, under the action of the operator $\mathbf{A},$ the image
vectors $\mathbf{A}\mathbf{u}_{i}$ span the area 
\[
\mathrm{area}(\mathbf{A}\mathbf{u}_{1},\ldots,\mathbf{\mathbf{A}}\mathbf{u}_{n-1})=\sqrt{\det\mathbf{G}(\mathbf{\mathbf{A}}\mathbf{u}_{1,}\ldots,\mathbf{A}\mathbf{u}_{n-1})}.
\]
Now, the volume of the $n$-dimensional parallelepiped formed by the
vectors, $n_{0},u_{1},\ldots,u_{n-1}$ is 
\[
\mathrm{vol}\left(\mathbf{n}_{0},\mathbf{u}_{1},\ldots,\mathbf{u}_{n-1}\right)=\sqrt{\det\mathbf{G}(\mathbf{u}_{1,}\ldots,\mathbf{u}_{n-1})},
\]
and hence the image of this parallelepiped under $\mathbf{A}$ has
the oriented volume 
\begin{align}
\mathrm{vol}\left(\mathbf{An}_{0},\mathbf{Au}_{1},\ldots,\mathbf{Au}_{n-1}\right) & =\det\mathbf{A}\,\mathrm{vol}\left(\mathbf{n}_{0},\mathbf{u}_{1},\ldots,\mathbf{u}_{n-1}\right)\label{eq:vol Au}\\
 & =\det\mathbf{A}\sqrt{\det\mathbf{G}(\mathbf{u}_{1,}\ldots,\mathbf{u}_{n-1})}.
\end{align}
With the unit normal $\mathbf{n}=\mathbf{A}^{-\top}\mathbf{n}_{0}/\left|\mathbf{A}^{-T}\mathbf{n}_{0}\right|$
to the image hyperplane $\mathbf{A}(E),$ we can also write 
\begin{align}
\mathrm{vol}\left(\mathbf{An}_{0},\mathbf{Au}_{1},\ldots,\mathbf{Au}_{n-1}\right) & =\left\langle \mathbf{A}\mathbf{n}_{0},\mathbf{n}\right\rangle \,\mathrm{area}(\mathbf{A}\mathbf{u}_{1},\ldots,\mathbf{\mathbf{A}}\mathbf{u}_{n-1})\nonumber \\
 & =\frac{\sqrt{\det\mathbf{G}(\mathbf{\mathbf{A}}\mathbf{u}_{1,}\ldots,\mathbf{A}\mathbf{u}_{n-1})}}{\sqrt{\left\langle \mathbf{n}_{0},\left(\mathbf{A}^{T}\mathbf{A}\right)^{-1}\mathbf{n}_{0}\right\rangle }}.\label{eq:vol Au2}
\end{align}
Therefore, a comparison of \eqref{eq:vol Au} and \eqref{eq:vol Au2}
gives 
\begin{equation}
\left\langle \mathbf{n}_{0},\left(\mathbf{A}^{\top}\mathbf{A}\right)^{-1}\mathbf{n}_{0}\right\rangle =\frac{\det\mathbf{G}(\mathbf{\mathbf{A}}\mathbf{u}_{1,}\ldots,\mathbf{A}\mathbf{u}_{n-1})}{\left(\det\mathbf{A}\right)^{2}\,\det\mathbf{G}(\mathbf{u}_{1,}\ldots,\mathbf{u}_{n-1})}.\label{eq:A formula}
\end{equation}

Back to the integral \eqref{eq: transport S21}, we note that the
symmetric tensor $\mathbf{\bar{T}}_{t_{0}}^{t_{1}}$ is positive definite,
and hence its inverse admits a unique symmetric, positive definite
square root tensor that can be written as $\left(\mathbf{\bar{T}}_{t_{0}}^{t_{1}}\right)^{-1}=\left(\mathbf{\bar{T}}_{t_{0}}^{t_{1}}\right)^{-\frac{1}{2}}\left(\mathbf{\bar{T}}_{t_{0}}^{t_{1}}\right)^{-\frac{1}{2}}$.
Then, selecting $\mathbf{A}=\left(\mathbf{\bar{T}}_{t_{0}}^{t_{1}}\right)^{-\frac{1}{2}}$
and $\mathbf{u}_{i}=\partial_{s_{i}}\mathbf{x}_{0}(\mathbf{s})$ in
formula \eqref{eq:A formula}, we conclude that the integral in \eqref{eq: transport S21}
can be re-written as{\footnotesize{}{} 
\[
\int_{\mathcal{M}_{0}}\left[\frac{\det\mathbf{\bar{T}}_{t_{0}}^{t_{1}}\,\det\mathbb{{\bf G}}\left(\left(\mathbf{\bar{T}}_{t_{0}}^{t_{1}}\right)^{-\frac{1}{2}}\partial_{\mathbf{s}}\mathbf{x}_{0}\right)}{\det\mathbf{G}}-\mathcal{T}\right]\sqrt{\det\mathbf{G}}\,ds_{1}\ldots ds_{n-1},
\]
}which proves the final formula we have given for $\mathcal{E}_{\mathcal{T}}$
with the Lagrangian $L$, as claimed.

\subsection*{S4: First integral and existence of uniform barriers}

The Lagrangian $L\left(\mathbf{x}_{0},\partial_{\mathbf{s}}\mathbf{x}_{0}\right)$
has no explicit dependence of the independent variable \textbf{$\mathbf{s}$},
and hence Noether's theorem provides partial conservation laws (cf.,
\cite{logan77}, Chapter 4, Example 4.2) for the associated Euler\textendash Lagrange
equation in the form 
\begin{equation}
\frac{\partial H_{j}^{i}}{\partial s_{k}}=0,\quad H_{j}^{i}:=\partial_{s_{j}}\mathbf{x}_{0}\cdot\frac{\partial L}{\partial\left(\partial_{s_{i}}\mathbf{x}_{0}\right)}-\delta_{ij}L,\quad i,j,k=1,\ldots,n-1,\label{eq:Hamiltonians}
\end{equation}
with $\delta_{ij}$ referring to the Kronecker delta. A direct calculation,
however, gives $H_{i}^{i}\equiv0,$ and hence no nontrivial conserved
quantity can be reconstructed from \eqref{eq:Hamiltonians}.

Instead, we apply an argument that extends the Maupertuis principle
derived for ordinary differential equations in \cite{moser03} to
partial differential equations. We start by considering another variational
problem associated with $\mathcal{E}_{\mathcal{T}_0}$ of the form 
\begin{equation}
\mathcal{\hat{E}}_{\mathcal{T}}=\int_{\kappa}G\left(\mathbf{x}_{0}(\mathbf{s}),\partial_{\mathbf{s}}\mathbf{x}_{0}(\mathbf{s})\right)\,d\mathbf{s,}\qquad G=L^{2}.\label{eq:newvari}
\end{equation}
As $G$ has no explicit dependence on $\mathbf{s}$, Noether's theorem
again applies and yields partial conservation laws given by \eqref{eq:Hamiltonians}.
In contrast to $L$, however, $G=L^{2}$ is a positively homogeneous
function of degree two, and hence, by Euler's theorem \cite{lewis69},
we obtain from \eqref{eq:Hamiltonians} for $i=j=k=1,\ldots,n-1$
that 
\[
\frac{\partial H_{i}^{i}}{\partial s_{i}}=0,\qquad H_{i}^{i}=\partial_{s_{i}}\mathbf{x}_{0}\cdot\frac{\partial G}{\partial\left(\partial_{s_{i}}\mathbf{x}_{0}\right)}-G=2G-G=G,
\]
and hence $L=\sqrt{G}$ is a first integral for the set of Euler\textendash Lagrange
partial differential equations 
\begin{equation}
G_{\mathbf{x}_{0}}-\sum_{i=1}^{n-1}\partial_{s_{i}}G_{\mathbf{x}_{0,i}}=\mathbf{0}.\label{eq:E-L G}
\end{equation}
(Here we have used the shorthand notation $G_{\mathbf{x}_{0}}:=\partial_{\mathbf{x}_{0}}G$
and $G_{\mathbf{x}_{0,i}}:=\partial_{\partial_{s_{i}}\mathbf{x}_{0}}G$.)
Consequently, 
\begin{equation}
G\left(\mathbf{x}_{0}(\mathbf{s}),\partial_{\mathbf{s}}\mathbf{x}_{0}(\mathbf{s})\right)=L^{2}\left(\mathbf{x}_{0}(\mathbf{s}),\partial_{\mathbf{s}}\mathbf{x}_{0}(\mathbf{s})\right)=const.\label{eq:G conserve}
\end{equation}
holds on the solutions $\mathbf{x}_{0}(\mathbf{s})$ of \eqref{eq:E-L G}.
We will now observe a close relationship between the solutions of
\eqref{eq:E-L G} and the solutions of the original variational problem.

To obtain this relationship, we first rewrite the left-hand side of
the Euler\textendash Lagrange equation 
\begin{equation}
\frac{\partial L}{\partial\mathbf{x}_{0}}-\sum_{i=1}^{n-1}\frac{\partial}{\partial s_{i}}\frac{\partial L}{\partial\left(\partial_{s_{i}}\mathbf{x}_{0}\right)}=\mathbf{0},\label{eq:E-L for L}
\end{equation}
for $L$ by substituting $L=\pm\sqrt{G}$, which gives 
\begin{align}
 & L_{\mathbf{x}_{0}}-\sum_{i=1}^{n-1}\partial_{s_{i}}L_{\mathbf{x}_{0,i}}\label{eq:E-L for G}\\
 & =\pm\frac{1}{2\sqrt{G}}\left[G_{\mathbf{x}_{0}}-\sum_{i=1}^{n-1}\partial_{s_{i}}G_{\mathbf{x}_{0,i}}\right]\mp\frac{\sum_{i=1}^{n-1}\partial_{s_{i}}G\, G_{\mathbf{x}_{0,i}}}{4\sqrt{G}^{3}},
\end{align}
whenever $G\neq0$. Therefore, a substitution of any solution solution
$\tilde{\mathbf{x}}_{0}(\mathbf{s})$ of the Euler\textendash Lagrange
\eqref{eq:E-L G} into \eqref{eq:E-L for G} gives 
\[
L_{\mathbf{x}_{0}}(\tilde{\mathbf{x}}_{0}(\mathbf{s}),\partial_{\mathbf{s}}\mathbf{\tilde{x}}_{0}(\mathbf{s}))-\sum_{i=1}^{n-1}\partial_{s_{i}}L_{\mathbf{x}_{0,i}}(\tilde{\mathbf{x}}_{0}(\mathbf{s}),\partial_{\mathbf{s}}\mathbf{\tilde{x}}_{0}(\mathbf{s}))=\mathbf{0},
\]
where we have used \eqref{eq:E-L G} and \eqref{eq:G conserve}. Therefore,
all solutions $\tilde{\mathbf{x}}_{0}(\mathbf{s})$ of \eqref{eq:E-L for G}
satisfying 
\begin{equation}
G(\tilde{\mathbf{x}}_{0}(\mathbf{s}),\partial_{\mathbf{s}}\mathbf{\tilde{x}}_{0}(\mathbf{s}))\neq0\label{eq: G not null}
\end{equation}
are also solutions on the Euler\textendash Lagrange \eqref{eq:E-L for L}.
Furthermore, since $G$ is constant along these solutions, $L=\pm\sqrt{G}$
is also constant along $\tilde{\mathbf{x}}_{0}(\mathbf{s})$.

Next, we assume that $\mathbf{x}_{0}(\mathbf{s})$ is a solution of
the Euler\textendash Lagrange \eqref{eq:E-L for L} for $L$. Rewriting
this equation using the relation $L=\pm\sqrt{G}$, we obtain{\footnotesize{}{}
\begin{equation}
\mathbf{0}=L_{\mathbf{x}_{0}}-\sum_{i=1}^{n-1}\partial_{s_{i}}L_{\mathbf{x}_{0,i}}=\pm\left[\frac{G_{\mathbf{x}_{0}}}{2\sqrt{G}}-\sum_{i=1}^{n-1}\partial_{s_{i}}\left(\frac{G_{\mathbf{x}_{0,i}}}{2\sqrt{G}}\right)\right].\label{eq:rescaled}
\end{equation}
}We now introduce a solution-dependent rescaling of the parameter
vector $\mathbf{s}$ by defining the new independent variable vector
$\mathbf{p}$ as 
\[
\mathbf{p}=\int_{\mathbf{s}_{0}}^{\mathbf{s}}{2\sqrt{G\left({\mathbf{x}}_{0}(\mathbf{\mathbf{\bm{\tau}}}),\partial_{\mathbf{s}}\mathbf{{x}}_{0}(\mathbf{\mathbf{\bm{\tau}}})\right)}\, d\mathbf{\bm{\tau}}},
\]
so that, in the new variable $\tilde{\mathbf{x}}_{0}(\mathbf{p})$, \eqref{eq:rescaled} becomes 
\begin{align*}
\mathbf{0} & =\pm\left[\frac{G_{\mathbf{x}_{0}}}{2\sqrt{G}}-\sum_{i=1}^{n-1}\partial_{s_{i}}\left(\frac{G_{\mathbf{x}_{0,i}}}{2\sqrt{G}}\right)\right]\\
 & =\pm {2\sqrt{G}}\left[G_{\tilde{\mathbf{x}}_{0}}\left(\tilde{\mathbf{x}}_{0},\partial_{\mathbf{p}}\mathbf{\tilde{x}}_{0}\right)-\sum_{i=1}^{n-1}\partial_{p_{i}}G_{\mathbf{\tilde{x}}_{0,i}}\left(\tilde{\mathbf{x}}_{0},\partial_{\mathbf{p}}\mathbf{\tilde{x}}_{0}\right)\right],
\end{align*}
where we have used the linearity of $G_{\mathbf{x}_{0,i}}$ in $\partial_{\mathbf{s}}\mathbf{x}_{0}$. Therefore, any solution $\mathbf{x}_{0}(\mathbf{s})$ of \eqref{eq:E-L for L}
satisfying \eqref{eq: G not null}, and hence satisfying $L(\tilde{\mathbf{x}}_{0}(\mathbf{s}),\partial_{\mathbf{s}}\mathbf{\tilde{x}}_{0}(\mathbf{s}))\neq0$,
is also a solution of \eqref{eq:E-L G} and thus conserves $G$, and
hence $L=\pm\sqrt{G}$, as first integrals. Consequently, all solutions
of \eqref{eq:E-L for L} and \eqref{eq:E-L G} are equivalent as long
as $L\neq0$ holds on them. This implies that the set $\mathcal{L}_{0}=\left\{ (\mathbf{x}_{0},\partial_{\mathbf{s}}\mathbf{x}_{0}):\,\,L(\mathbf{x}_{0},\partial_{\mathbf{s}}\mathbf{x}_{0})=0\right\} $,
if nonempty, is an invariant set for the Euler\textendash Lagrange
equation of $L$.

\subsection*{S5: Local necessary conditions for extrema}

If $\mathcal{M}_{0}$ is a stationary surface for a quotient functional
$Q=A/B$ with $B>0$, then we have 
\[
\delta Q\vert_{\mathcal{M}_{0}}=\frac{\delta\left(A-\frac{A_{0}}{B_{0}}B\right)}{B}\vert_{\mathcal{M}_{0}}=0,\quad\delta^{2}Q\vert_{\mathcal{M}_{0}}=\frac{\delta^{2}\left(A-\frac{A_{0}}{B_{0}}B\right)}{B}\vert_{\mathcal{M}_{0}},
\]
with $A_{0}:=A\vert_{\mathcal{M}_{0}}$ and $B_{0}:=B\vert_{\mathcal{M}_{0}}$.
Consequently, local maxima (or minima) of $\mathcal{T}{}_{t_{0}}^{t_{1}}$
coincide with the local maxima (or minima, respectively) of $\mathcal{E}_{\mathcal{T}_0}\left(\mathcal{M}_{0}\right)$.

A simple necessary condition for a null-surface $\mathcal{M}_{0}$
to be an extremizer of $\mathcal{E}_{\mathcal{T}_0}$ can be obtained
by considering a small, surface-area-preserving perturbation $\mathbf{x}_{0}^{\epsilon}(\mathbf{s})=\mathbf{x}_{0}(\mathbf{s})+\epsilon\mathbf{h}_{\epsilon}\left(\left(\mathbf{s}-\mathbf{s}_{0}\right)/\epsilon\right)$
to $\mathcal{M}_{0}$ , where $\mathbf{h}_{\epsilon}\colon\mathbb{R}^{n-1}\to\mathbb{R}^{n}$
is a uniformly bounded, smooth function with $\mathbf{h}_{\epsilon}(\mathbf{0})=\mathbf{0}$,
$D\mathbf{h}_{\epsilon}(\mathbf{0})\neq\mathbf{0}$ that is supported
only in an $\mathcal{O}(\epsilon)$ neighborhood of the origin. The
function $\mathbf{x}_{0}^{\epsilon}(\mathbf{s})$ then gives the parametrization
of a perturbed hypersurface $\mathcal{M}_{0}^{\epsilon}$. Within
the support of $\mathbf{h}$, the unit normal $\mathbf{n}_{0}^{\epsilon}(\mathbf{s}_{0})$
of the perturbed surface at $\mathbf{s}_{0}$ must therefore satisfy
\[
\mathbf{n}_{0}^{\epsilon}(\mathbf{s}_{0})=\left[1-\mathcal{O}(\epsilon)\right]\mathbf{n}_{0}^{\perp}(\mathbf{s}_{0})+\mathcal{O}(\epsilon),\quad\left\langle \mathbf{n}_{0}^{\perp}(\mathbf{s}_{0}),\mathbf{n}_{0}(\mathbf{s}_{0})\right\rangle =0
\]
for some $\left|\mathbf{n}_{0}^{\perp}\right|=1.$ For $\mathbf{s}$
values outside the support of $\mathbf{h}_{\epsilon}$, we have $\mathbf{n}_{0}^{\epsilon}(\mathbf{s})\equiv\mathbf{n}_{0}(\mathbf{s})$.
One then obtains 
\begin{align*}
\mathcal{E}_{\mathcal{T}_0}(\mathcal{M}_{0}^{\epsilon}) & =\int_{\mathcal{M}_{0}\cap\mathcal{M}_{0}^{\epsilon}}\left\langle \mathbf{n}_{0}^{\epsilon},\mathbf{\mathbf{E}_{\mathcal{T}_0}}\mathbf{n}_{0}^{\epsilon}\right\rangle dA_{0}\\
& +\int_{\mathcal{M}_{0}^{\epsilon}-\mathcal{M}_{0}}\left\langle \mathbf{n}_{0}^{\epsilon},\mathbf{\mathbf{E}_{\mathcal{T}_0}}\mathbf{n}_{0}^{\epsilon}\right\rangle dA_{0}
\end{align*}
\begin{equation*}
 =\int_{\mathcal{M}_{0}^{\epsilon}-\mathcal{M}_{0}}\left[\left\langle \mathbf{n}_{0}^{\perp}(\mathbf{s}_{0}),\mathbf{\mathbf{E}_{\mathcal{T}_0}}\left(\mathbf{x}_{0}(\mathbf{s}_{0})\right)\mathbf{n}_{0}^{\perp}(\mathbf{s}_{0})\right\rangle +\mathcal{O}(\epsilon)\right]dA_{0}
 \end{equation*}
\begin{equation*}
=\left\langle \mathbf{n}_{0}^{\perp}(\mathbf{s}_{0}),\mathbf{\mathbf{E}_{\mathcal{T}_0}}\left(\mathbf{x}_{0}(\mathbf{s}_{0})\right)\mathbf{n}_{0}^{\perp}(\mathbf{s}_{0})\right\rangle \mathrm{vol}_{n-1}\left(\mathcal{M}_{0}^{\epsilon}-\mathcal{M}_{0}\right)+\mathcal{O}(\epsilon^{n}),
\end{equation*}
where we have used that $\left\langle \mathbf{n}_{0}^{\epsilon},\mathbf{\mathbf{E}_{\mathcal{T}_0}}\mathbf{n}_{0}^{\epsilon}\right\rangle =\left\langle \mathbf{n}_{0},\mathbf{\mathbf{E}_{\mathcal{T}_0}}\mathbf{n}_{0}\right\rangle =0$
holds along $\mathcal{M}_{0}$, and that the support of $\mathbf{h}_{\epsilon}$
has volume of order $\mathcal{O}(\epsilon^{n-1})$ in $\mathbb{R}^{n-1}$.
Therefore, if $\mathcal{M}_{0}$ is a local minimizer of the functional
$\mathcal{E}_{\mathcal{T}_0}$, then we must necessarily have 
\[
\left\langle \mathbf{n}_{0}^{\perp}(\mathbf{s}_{0}),\mathbf{\mathbf{E}_{\mathcal{T}_0}}\left(\mathbf{x}_{0}(\mathbf{s}_{0})\right)\mathbf{n}_{0}^{\perp}(\mathbf{s}_{0})\right\rangle \geq0.
\]
Since the point $\mathbf{x}_{0}(\mathbf{s}_{0})$ along $\mathcal{M}_{0}$
and the exact shape of $\mathbf{h}_{\epsilon}$ (and hence $\mathbf{n}_{0}^{\perp}(\mathbf{s}_{0})\in T_{\mathbf{x}_{0}}\mathcal{M}_{0})$
have been arbitrary, this last inequality implies 
\begin{equation}
\left\langle \mathbf{u},\mathbf{\mathbf{E}_{\mathcal{T}_0}}\left(\mathbf{x}_{0}\right)\mathbf{u}\right\rangle \geq0,\qquad\forall\mathbf{u}\in T_{\mathbf{x}_{0}}\mathcal{M}_{0},\quad\forall\mathbf{x}_{0}\in\mathcal{M}_{0}.\label{eq:pos semidef on bundle}
\end{equation}
Therefore, the tensor $\mathbf{\mathbf{E}_{\mathcal{T}_0}}$ must be
positive semidefinite on the tangent bundle $T\mathcal{M}_{0}$ of
its null surface $\mathcal{M}_{0}$, if this null surface is a transport
barrier.

Next, we derive a condition equivalent to \eqref{eq:pos semidef on bundle}
that is nevertheless easier to verify directly from the eigenvalues
of $\mathbf{\mathbf{E}_{\mathcal{T}_0}}\left(\mathbf{x}_{0}\right)$.
To this end, let us denote the eigenvalues of $\mathbf{\mathbf{E}_{\mathcal{T}_0}}\left(\mathbf{x}_{0}\right)$
by 
\[
\rho_{1}\left(\mathbf{x}_{0}\right):=\mu_{1}(\mathbf{x}_{0})-\mathcal{T}_0\leq\ldots\leq\rho_{n}\left(\mathbf{x}_{0}\right):=\mu_{n}(\mathbf{x}_{0})-\mathcal{T}_0,
\]
with $0<\mu_{1}(\mathbf{x}_{0})\leq\ldots\leq\mu_{n}(\mathbf{x}_{0})$
denoting the eigenvalues of the positive definite tensor $\mathbf{\bar{T}}_{t_{0}}^{t_{1}}$,
as earlier. We observe that condition \eqref{eq:pos semidef on bundle}
implies $\mathbf{\rho}_{1}(\mathbf{x}_{0})\leq0\leq\mathbf{\rho}_{n}(\mathbf{x}_{0}).$
Indeed, if $\mathbf{\rho}_{1}(\mathbf{x}_{0})>0$ or $\mathbf{\rho}_{n}(\mathbf{x}_{0})<0$
were satisfied, then $\mathbf{\mathbf{E}_{\mathcal{T}_0}}\left(\mathbf{x}_{0}\right)$
would be definite and hence could have no nonempty null-surface $\mathcal{M}_{0}$.

We next show that 
\begin{equation}
\mathbf{\rho}_{k}(\mathbf{x}_{0})\geq0,\qquad k\geq2,\label{eq:mu_k > 0}
\end{equation}
must necessarily hold. Indeed, assuming the opposite would imply,
by the ordering of the eigenvalues, that $\mathbf{\rho}_{2}(\mathbf{x}_{0})<0$
holds, and hence $\mathbf{\mathbf{E}_{\mathcal{T}_0}}$ would have two
negative eigenvalues, $\mathbf{\rho}_{1}(\mathbf{x}_{0})$ and $\mathbf{\rho}_{2}(\mathbf{x}_{0}).$
This would then necessarily imply that $\mathbf{\rho}_{n}(\mathbf{x}_{0})>0$
(otherwise the unit normal $\mathbf{n}_{0}(\mathbf{x}_{0})$ would
necessarily have to be orthogonal to the eigenvectors of these two
negative eigenvalues, and $\left\langle \mathbf{u},\mathbf{\mathbf{E}_{\mathcal{T}_0}}\left(\mathbf{x}_{0}\right)\mathbf{u}\right\rangle $
would necessarily take negative values in $T_{\mathbf{x}_{0}}\mathcal{M}_{0}$).
Therefore, \eqref{eq:mu_k > 0} must be satisfied.

Finally, we show that 
\begin{equation}
-\mathbf{\rho}_{1}(\mathbf{x}_{0})\leq\mathbf{\rho}_{n}(\mathbf{x}_{0})\label{eq:finalcond}
\end{equation}
must hold. Indeed, assuming $-\mathbf{\rho}_{1}(\mathbf{x}_{0})>\mathbf{\rho}_{n}(\mathbf{x}_{0})$
necessarily implies $\mathbf{\rho}_{1}(\mathbf{x}_{0})<0<\mathbf{\rho}_{n}(\mathbf{x}_{0})$
must hold, and hence, by \eqref{eq:mu_k > 0}, the local unit normal
$\mathbf{n}_{0}=\left(n_{01},\ldots,n_{0n}\right)$ of $\mathcal{M}_{0}$,
with coordinates with respect to the orthonormal eigenbasis $\left\{ \mbox{\ensuremath{\bm{\zeta}}}_{1},\ldots\bm{\zeta}_{n}\right\} $
of $\mathbf{\mathbf{E}_{\mathcal{T}_0}}$, must satisfy the equation
\begin{equation}
n_{01}^{2}=\frac{\rho_{2}}{\left|\rho_{1}\right|}n_{02}^{2}+\ldots+\frac{\rho_{n}}{\left|\rho_{1}\right|}n_{0n}^{2},\label{eq:n surface}
\end{equation}
where all coefficients on the right-hand side are nonnegative, and
at least $\frac{\rho_{n}}{\left|\rho_{1}\right|}$ is strictly positive.
The surface $\mathcal{C}$ defined by \eqref{eq:n surface} is a codimension-one
elliptical cone when the coefficients $\rho_{2},\ldots,\rho_{n}$
are nonzero, or the product of a lower-dimensional elliptical cone
with a plane when some of these coefficients are zero. Consider now
a codimension-one plane $\mathcal{P}$ containing the normal $\mathbf{n}_{0}$
and the $\mbox{\ensuremath{\bm{\zeta}}}_{1}$ axis. The intersection
$\mathcal{C}\cap\mathcal{P}$ then consists of two lines, one through
$\mathbf{n}_{0}$ and another line through the mirror image $\hat{\mathbf{n}}_{0}=2\left\langle \mathbf{n}_{0},\mbox{\ensuremath{\bm{\zeta}}}_{1}\right\rangle \mbox{\ensuremath{\bm{\zeta}}}_{1}-\mathbf{n}_{0}$
of $\mathbf{n}_{0}$ with respect to the $\mbox{\ensuremath{\bm{\zeta}}}_{1}$
axis. If the angle of $\mathbf{n}_{0}$ and $\hat{\mathbf{n}}_{0}$
is more than $\pi/2$ than then the plane normal to $\mathbf{n}_{0}$
also intersects $\mathcal{C}$ transversely, and hence $\left\langle \mathbf{u},\mathbf{\mathbf{E}_{\mathcal{T}_0}}\left(\mathbf{x}_{0}\right)\mathbf{u}\right\rangle $
will change its sign within the tangent plane $T_{\mathbf{x}_{0}}\mathcal{M}_{0}.$
Consequently, the minimal possible angle between $\mathbf{n}_{0}$
and $\hat{\mathbf{n}}_{0}$, over all choices of $\mathbf{n}_{0}$
at a point $\mathbf{x}_{0}\in\mathcal{M}_{0}$, cannot exceed $\pi/2$,
otherwise $\mathcal{M}_{0}$ cannot be a diffusive transport minimizer.
This minimal angle arises when $\mathbf{n}_{0}$ is contained in the
subspace of the elliptical cone $\mathcal{C}$ that runs closest to
the $\mbox{\ensuremath{\bm{\zeta}}}_{1}$ axis, i..e, when $n_{02}=\ldots=n_{0(n-1)}$
are zero. In this case, $n_{01}\pm=\sqrt{\frac{\rho_{n}}{\left|\rho_{1}\right|}n_{0n}^{2}}$,
and hence the angle between $\mathbf{n}_{0}$ and $\hat{\mathbf{n}}_{0}$
exceeds $\pi/2$, given that we have assumed $-\mathbf{\rho}_{1}(\mathbf{x}_{0})=\left|\rho_{1}\right|>\mathbf{\rho}_{n}(\mathbf{x}_{0})$.
We, therefore, conclude that \eqref{eq:finalcond} must hold.

In summary, the inequalities \eqref{eq:mu_k > 0} and \eqref{eq:finalcond}give
the necessary conditions $\mu_{k}(\mathbf{x}_{0})\geq\mathcal{T}_0,k=2,\ldots,n-1$,
and $\mathbf{\mu}_{n}(\mathbf{x}_{0})-\mathcal{T}_0\geq\mathcal{T}_0-\mathbf{\mu}_{1}(\mathbf{x}_{0})$.
Summing up these inequalities then gives the necessary condition $0\leq\mu_{1}(\mathbf{x}_{0})+\ldots+\mathbf{\mu}_{n}(\mathbf{x}_{0})-n\mathcal{T}_0=\mathrm{trace\,}\mathbf{\mathbf{E}_{\mathcal{T}_0}}\left(\mathbf{x}_{0}\right)$
for transport barriers, as claimed. A similar argument applied to
transport enhancers gives then the necessary condition $\mathrm{trace\,}\mathbf{\mathbf{E}_{\mathcal{T}_0}}\left(\mathbf{x}_{0}\right)\leq0$.

\subsection*{S6: Transport extremizers in two dimensions}

We first introduce the diffusion-weighted Cauchy\textendash Green
strain tensor $\mathbf{C}_{\mathbf{D}}:=\det\left[\mathbf{D}\left(\mathbf{F}_{t_{0}}^{t},t\right)\right]\left[\mathbf{T}_{t_{0}}^{t}\right]^{-1}.$
Denoting by $\left(\mathbf{T}_{t_{0}}^{t}\right)^{c}$ the co-factor
matrix of $\mathbf{T}_{t_{0}}^{t}$, we observe that by incompressibility
($\det\bm{\nabla}_{0}\mathbf{F}_{t_{0}}^{t}\equiv1$), we have 
\begin{align}
\left(\mathbf{T}_{t_{0}}^{t}\right)^{c} & =\left(\mathbf{T}_{t_{0}}^{t}\right)^{-1}/\det\left(\mathbf{T}_{t_{0}}^{t}\right)^{-1}=\mathbf{C}_{\mathbf{D}}.\label{eq:cofactor ID}
\end{align}
We further note that in case of homogeneous-isotropic diffusion ($\mathbf{D}\equiv\mathbf{I}$),
we have $\det\left(\mathbf{T}_{t_{0}}^{t}\right)^{-1}\equiv1$, and
hence \eqref{eq:cofactor ID} gives 
\begin{align}
\mathbf{\bar{T}}_{t_{0}}^{t_{1}} & =\frac{1}{t_{1}-t_{0}}\int_{t_{0}}^{t_{1}}\mathbf{T}_{t_{0}}^{t}dt=\frac{1}{t_{1}-t_{0}}\int_{t_{0}}^{t_{1}}\mathbf{C}_{\mathbf{D}}^{-1}dt\nonumber \\
 & =\frac{1}{t_{1}-t_{0}}\int_{t_{0}}^{t_{1}}\left(\mathbf{C}_{\mathbf{D}}\right)^{c}dt=\left(\frac{1}{t_{1}-t_{0}}\int_{t_{0}}^{t_{1}}\mathbf{C}_{\mathbf{D}}dt\right)^{c}\nonumber \\
 & =\left(\det\mathbf{\bar{C}}_{\mathbf{D}}\right)\mathbf{\bar{C}}_{\mathbf{D}}^{-1},\label{eq:TDformula1}
\end{align}
which further implies 
\begin{equation}
\mu_{i}=\lambda_{i},\quad\bm{\zeta}_{i}=\bm{\xi}_{j},\quad i,j=1,2,\quad i\neq j,\label{eq:TDformula2}
\end{equation}
where $0<\mu_{1}\leq\mu_{2}$ denote the eigenvalues of $\mathbf{\bar{T}}_{t_{0}}^{t_{1}}$
corresponding to the orthonormal eigenbasis $\left\{ \bm{\zeta}_{1},\bm{\zeta}_{2}\right\} $
and $\bm{\xi}_{j}$ denote the normalized eigenvectors of $\mathbf{\bar{C}}_{\mathbf{D}}$.

Using \eqref{eq:TDformula1}, we obtain the Lagrangian $L$ for two-dimensional
flows in the form 
\begin{align*}
L & =\frac{\,\left\langle \sqrt{\det\mathbf{\bar{T}}_{t_{0}}^{t_{1}}}\left(\mathbf{\bar{T}}_{t_{0}}^{t_{1}}\right)^{-\frac{1}{2}}\mathbf{x}_{0}^{\prime},\sqrt{\det\mathbf{\bar{T}}_{t_{0}}^{t_{1}}}\left(\mathbf{\bar{T}}_{t_{0}}^{t_{1}}\right)^{-\frac{1}{2}}\mathbf{x}_{0}^{\prime}\right\rangle }{\sqrt{\left\langle \mathbf{x}_{0}^{\prime}(s),\mathbf{x}_{0}^{\prime}(s)\right\rangle }}\\
 & -\mathcal{T}_0\sqrt{\left\langle \mathbf{x}_{0}^{\prime},\mathbf{x}_{0}^{\prime}\right\rangle }\\
 & =\frac{1}{t_{1}-t_{0}}\int_{t_{0}}^{t_{1}}\frac{\,\left\langle \mathbf{x}_{0}^{\prime},\left(\mathbf{T}_{t_{0}}^{t_{1}}\right)^{c}\mathbf{x}_{0}^{\prime}\right\rangle }{\sqrt{\left\langle \mathbf{x}_{0}^{\prime},\mathbf{x}_{0}^{\prime}\right\rangle }}dt-\mathcal{T}_0\sqrt{\left\langle \mathbf{x}_{0}^{\prime},\mathbf{x}_{0}^{\prime}\right\rangle },
\end{align*}
which, together with \eqref{eq:cofactor ID}, gives 
\[
L=\frac{\,\left\langle \mathbf{x}_{0}^{\prime},\mathbf{\bar{C}}_{\mathbf{D}}(\mathbf{x}_{0})\mathbf{x}_{0}^{\prime}\right\rangle }{\sqrt{\left\langle \mathbf{x}_{0}^{\prime},\mathbf{x}_{0}^{\prime}\right\rangle }}-\mathcal{T}_0\sqrt{\left\langle \mathbf{x}_{0}^{\prime},\mathbf{x}_{0}^{\prime}\right\rangle }=\frac{C_{ij}w_{i}w_{j}}{\sqrt{w_{k}w_{k}}}-\mathcal{T}_0\sqrt{w_{k}w_{k}},
\]
with the simplified notation $\mathbf{x}=\mathbf{x}_{0},$ $\mathbf{w}=\mathbf{x}_{0}^{\prime}$
, $\mathbf{a}=\mathbf{x}_{0}^{\prime\prime}$ , and $\mathbf{C}=\bar{\mathbf{C}}_{\mathbf{D}}$.
From this, we obtain the Euler\textendash Lagrange equations $L_{\mathbf{x}_{0}}-\frac{d}{ds}L_{\mathbf{x}_{0}^{\prime}}=0$
for $L$ in coordinate form as{\footnotesize{}{} 
\begin{align*}
\left(\mathcal{T}_0\left|\mathbf{w}\right|^{2}+C_{ij}v_{i}v_{j}\right)a_{m}\\
-\left[2\left|\mathbf{w}\right|^{2}C_{mk}-2C_{mj}v_{j}v_{k}-2C_{kj}v_{j}v_{m}+3\frac{\frac{1}{3}\mathcal{T}_0\left|\mathbf{w}\right|^{2}+C_{ij}v_{i}v_{j}}{\left|\mathbf{w}\right|^{2}}v_{m}v_{k}\right]a_{k}\\
+C_{ij,l}v_{i}v_{j}v_{l}v_{m}+\left(C_{ij,m}v_{i}v_{j}-2C_{mj,l}v_{j}v_{l}\right)\left|\mathbf{w}\right|^{2}=0 & .
\end{align*}
}{\footnotesize \par}

Recall that the boundary term arising in the conversion of the weak
form of Euler\textendash Lagrange equation to its strong form must
vanish, which gives, in two dimensions, the requirement 
\[
\partial_{\mathbf{x}_{0}^{\prime}}\left[\frac{\left\langle \mathbf{x}_{0}^{\prime},\bar{\mathbf{C}}_{\mathbf{D}}(\mathbf{x}_{0})\mathbf{x}_{0}^{\prime}\right\rangle }{\mathcal{}\sqrt{\left\langle \mathbf{x}_{0}^{\prime},\mathbf{x}_{0}^{\prime}\right\rangle }}-\mathcal{T}_0\sqrt{\left\langle \mathbf{x}_{0}^{\prime},\mathbf{x}_{0}^{\prime}\right\rangle }\right]_{\partial\mathcal{M}_{0}}\cdot\mathbf{h}=0,
\]
with $\partial\mathcal{M}_{0}$ denoting just a pair of discrete points.
Evaluating this condition along uniform extremizers and noting the
relations $\left|\mathbf{\bm{\eta}_{\mathcal{T}_0}}\right|=1$ and $\mathbf{x}_{0}^{\prime}\parallel\bm{\eta}_{\mathcal{T}_0}(\mathbf{x}_{0})\perp\mathbf{h}(\mathbf{x}_{0})$
at $\mathbf{x}_{0}\in\partial\mathcal{M}_{0}$, we obtain 
\begin{align*}
 & \left[2\bar{\mathbf{C}}_{\mathbf{D}}\mathbf{\bm{\eta}_{\mathcal{T}_0}}-2\mathcal{T}_0\mathbf{\mathbf{\bm{\eta}_{\mathcal{T}_0}}}\right]\cdot\mathbf{h}=2\left\langle \mathbf{\hat{E}}_{\mathcal{T}_0}\mathbf{\bm{\eta}_{\mathcal{T}_0}},\mathbf{h}\right\rangle =0.
\end{align*}
This inner product only vanishes in the following three cases: 
\begin{description}
\item [{($B1_{2D}$)}] \emph{Normal boundary perturbations} (\textbf{front-type
surfaces}; $\hat{\mathbf{E}}_{\mathcal{T}_0}\mathbf{\bm{\eta}_{\mathcal{T}_0}}\perp\mathbf{h})$
This is only possible at a boundary point $\mathbf{x}_{0}\in\partial\mathcal{M}_{0}$
if $\mathbf{\bm{\eta}_{\mathcal{T}_0}}(\mathbf{x}_{0})$ is an eigenvector
of $\mathbf{\hat{E}}_{\mathcal{T}_0}(\mathbf{x}_{0})$, i.e., $\mathbf{\bm{\eta}_{\mathcal{T}_0}}(\mathbf{x}_{0})=\bm{\xi}_{i}(\mathbf{x}_{0})$
holds for some $i\in\left\{ 1,2\right\} ,$ with $\bm{\xi}_{i}$ denoting
the unit eigenvectors of the tensor $\bar{\mathbf{C}}_{\mathbf{D}}$.
This condition holds at maximal open null-geodesics of $\mathbf{\hat{E}}_{\mathcal{T}_0}(\mathbf{x}_{0})$,
i.e., $\mathbf{\bm{\eta}_{\mathcal{T}_0}}$-lines ending at points where
$\mathbf{\hat{E}}_{\mathcal{T}_0}(\mathbf{x}_{0})$ has precisely one
zero eigenvalue. 
\item [{($B2_{2D}$)}] \emph{Boundary perturbations along a two-dimensional
subspace (}\textbf{jet-core-type surfaces}\emph{; $\mathbf{\hat{E}}_{\mathcal{T}_0}(\mathbf{x}_{0})=\mathbf{0}$
).} This is only possible at a boundary point $\mathbf{x}_{0}\in\partial\mathcal{M}_{0}$
if the symmetric tensor $\mathbf{\hat{E}}_{\mathcal{T}_0}(\mathbf{x}_{0})$
has two zero eigenvalues. That happens precisely when $\bar{\mathbf{C}}_{\mathbf{D}}(\mathbf{x}_{0})$
has two repeated eigenvalues satisfying $\lambda_{1}(\mathbf{x}_{0})=\lambda_{2}(\mathbf{x}_{0})=\mathcal{T}_0.$ 
\item [{($B3_{2D}$)}] \emph{Empty boundary} (\textbf{closed vortical surfaces};
$\partial\mathcal{M}_{0}=\emptyset)$: Such extremizers have no boundaries
and hence are closed $\mathbf{\bm{\eta}_{\mathcal{T}_0}}$-lines (limit
cycles) of the direction field $\mathbf{\bm{\eta}_{\mathcal{T}_0}}(\mathbf{x}_{0})$. 
\end{description}

\subsection*{S7: Numerical algorithm in two dimensions and description of the
examples}

We have summarized the main steps in the computation of diffusion
barriers in steps (A1)-(A5). A fundamental requirement in these steps
is the accurate computation of the eigenvalues and eigenvectors of
the tensor field $\mathbf{\bar{C}}_{\mathbf{D}}(\mathbf{x}_{0}).$
The numerical challenges involved in this computation are identical
to those faced in computing the right Cauchy\textendash Green strain tensor
$\mathbf{C}_{t_{0}}^{t}(\mathbf{x}_{0})$, as discussed in \cite{haller15}.

Closed diffusion barriers can be computed by finding outermost limit
cycles of $\mathbf{\bm{\eta}_{\mathcal{T}_{0}}}(\mathbf{x}_{0})$ that we carry out using a modification of the algorithm used in \cite{hadjighasem16c},
which is originally based on \cite{karrasch15}. These modifications
include improvements in determining singularity types for the direction
field $\mathbf{\bm{\eta}_{\mathcal{T}_{0}}}(\mathbf{x}_{0})$, as well
as refinements to finding zeros of Poincaré maps that capture limit
cycles of this field.

For two-dimensional flows, the cost of closed diffusive and
stochastic barrier computations is close to that of the computations
of elliptic deterministic transport barriers (geodesic LCS) for deterministic
flows, because Eq. (22) is formally identical to that defining elliptic
LCSs \cite{haller13}. The only difference is that the eigenvalues and eigenvectors
appearing in Eq. (22) are those of $\mathbf{\bar{C}}_{\mathbf{D}}$,
as opposed to those of $\mathbf{C}_{t_{0}}^{t}$
in the deterministic case \cite{haller13}. The temporal averaging of $\mathbf{\bar{C}}_{\mathbf{D}}$
practically requires the computation of $\mathbf{C}_{t_{0}}^{t}$
at intermediate times, not just at the final time, as in geodesic
LCS theory. This, however, adds a negligible increment in computation
times, as the most time-consuming part of the algorithm (advection
of an initial grid) is the same in both cases. For the same reason,
the cost of computing the DBS diagnostic field for hyperbolic and
parabolic diffusion barriers is practically identical to that of the
finite-time Lyapunov exponent (FTLE) field used in the deterministic
setting \cite{haller15}.

Diffusive barriers, therefore, differ from their deterministic
counterparts (LCSs) because of the appearance of the diffusion structure
tensor and temporal averaging in the computation of the tensor $\mathbf{\bar{C}}_{\mathbf{D}}$.
For small diffusivities, this mismatch is independent of the value
of the diffusivity and will be larger when the diffusion structure
tensor $\mathbf{D}$ is far from the identity tensor, or when the
averaged Cauchy\textendash Green strain tensor $\mathbf{\bar{C}}_{\mathbf{D}}$
is far from its unaveraged counterpart. The former case arises under
significant anisotropy in the diffusion, while the latter case arises
under significant temporal aperiodicity in the velocity field.

To solve the time-dependent advection-diffusion equation in two-dimensions,
we use a finite-element (FE) discretization in space, and employ an implicit
Euler time-stepping scheme with fixed stepsize. Our FEM implementation is
based on JuAFEM, a simple finite element toolbox written in Julia.

As for our stochastic formulation involving Eq. (23), we change the physical (Eulerian) coordinate $\mathbf{x}$ of fluid
trajectories to their initial conditions $\mathbf{x}_{0}$ for our simulations. This is done through
the deterministic relationship $\mathbf{x}=\mathbf{F}_{t_{0}}^{t}(\mathbf{x}_{0})$,
which yields $d\mathbf{x}(t)=\bm{\nabla}_{0}\mathbf{F}_{t_{0}}^{t}\left(\mathbf{x}_{0}(t)\right)d\mathbf{x}_{0}(t)+\frac{\partial}{\partial t}\mathbf{F}_{t_{0}}^{t}\left(\mathbf{x}_{0}(t)\right)dt$.
Comparing this differential with the stochastic differential Eq. 
(23) then yields the Lagrangian form of Eq. 
(23) we have given in Section 7 (see \cite{fyrillas07} for an earlier derivation). The time-dependence in the
Lagrangian variable $\mathbf{x}_{0}(t)$ is solely due to the presence
of the Brownian motion in (23), which turns the initial condition
obtained through the deterministic relationship $\mathbf{x}_{0}=\mathbf{F}_{t}^{t_{0}}(\mathbf{x})$
into a stochastic, time-dependent variable.

To simulate trajectories of Eq. (27), we
first compute the pullback diffusion matrix field as
\[
\mathbf{B}_0(\mathbf{x}_0,t)=\left[\mathbf{\bm{\nabla}}_{0}\mathbf{F}_{t_{0}}^{t}(\mathbf{x}_0)\right]^{-1}\mathbf{B}\left(\mathbf{F}_{t_{0}}^{t}(\mathbf{x}_0),t\right),
\]
using the deformation gradient $\bm{\nabla}_{0}\mathbf{F}_{t_{0}}^{t}$ computed as above. Subsequently, the matrix field is interpolated in space-time, and the stochastic trajectories of
\[
d\mathbf{x}_0(t)=\sqrt{\nu}\mathbf{B}_0(\mathbf{x}_0(t),t)d\mathbf{W}(t)
\]
are then computed using Rössler's adaptive strong order 1.5 method \cite{roessler10}, as implemented in the StochasticDiffEq.jl package of Julia. We release 50
trajectories per initial condition, arranged in a coarser uniform grid; see the
animation in \emph{SI Appendix S9} for the initial configuration.

\begin{figure}[b]
\centering
\includegraphics[width=1\linewidth]{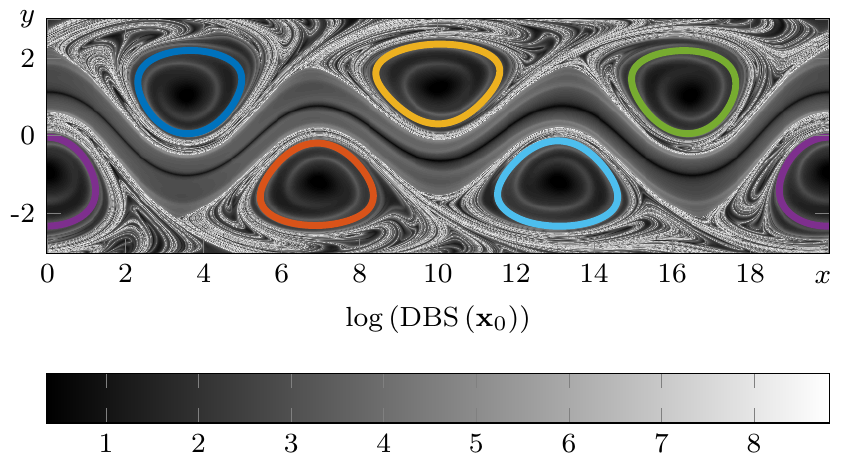}
\caption{Diffusion barriers in the Bickley jet: vortex boundaries (outermost
limit cycles of the $\mathbf{\bm{\eta}_{\mathcal{T}_{0}}}(\mathbf{x}_{0})$
field), backward-fronts (ridges of the $\DBS(\mathbf{x}_{0})$
field), and a jet core (trench of the $\DBS(\mathbf{x}_{0})$
field). See \emph{Supporting Animation SA2} for the evolution of the advected vortex boundaries.}
\label{fig:bickley-chi} 
\end{figure}

\begin{figure}
\centering
\includegraphics[width=1\linewidth]{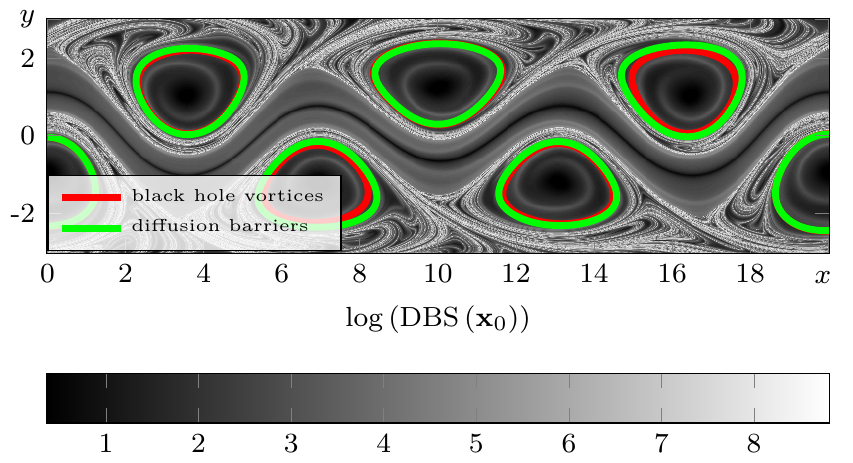}
\caption{Red: Elliptic LCSs (black-hole vortex boundaries computed from the
theory in \cite{haller13} for the Bickley jet. Green: outermost closed
diffusion barriers computed from the present theory. Background: the
$\chi({\mathbf{x}}_{0})$ field.}
\label{bickley-BH-vs-diff} 
\end{figure}
\begin{figure}
\centering
\includegraphics[width=1\linewidth]{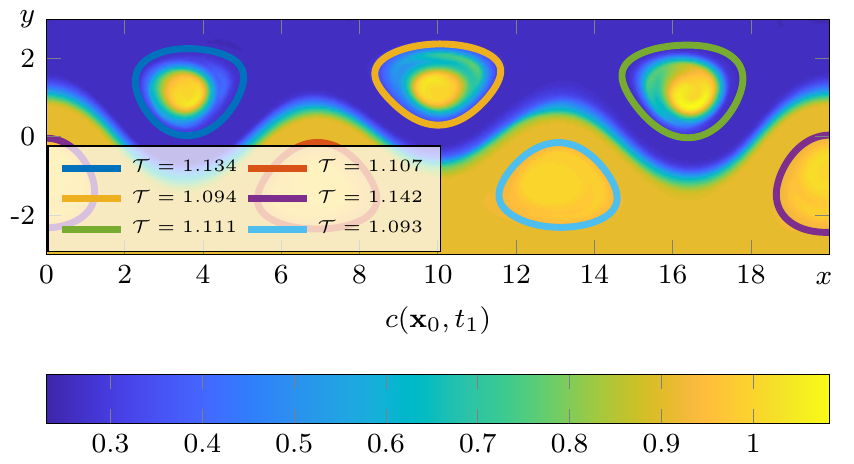}
\caption{Diffused distribution of $c(\mathbf{x}_{0},t_{1})$, the tracer field
in Lagrangian coordinates at time $t_{1}=40$ days for the Bickley
jet. The initial tracer distribution $c_{0}(\mathbf{x}_{0})$ was
selected constant and unity inside the diffusion barriers encircling
the upper vortices, as well as below the diffusion barrier acting
as the jet core.}
\label{fig:bickley-diff} 
\end{figure}
\begin{figure}
\centering
\includegraphics{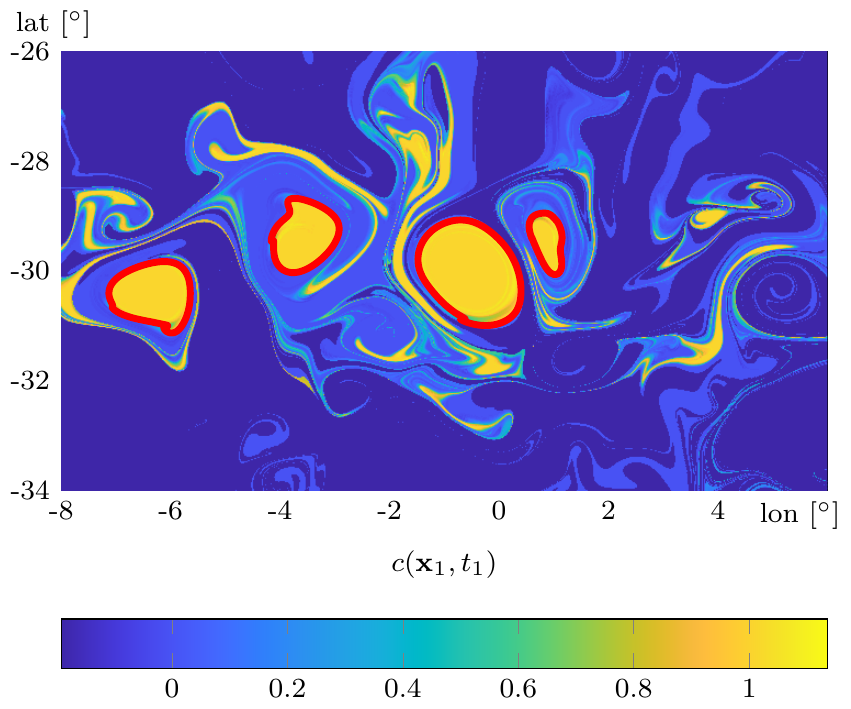}
\caption{Diffused concentration $c(\mathbf{x}_1,t_1)$ at time $t_1$ in Eulerian coordinates $\mathbf{x}_1 = \mathbf{F}_{t_0}^{t_1}(\mathbf{x}_0)$, with the advected position of diffusion-based ring boundaries overlaid. The initial concentration was localized on the four coherent vortices and seven shifted copies, cf.~Fig.~3 in the main text. See also \emph{Supporting Animation SA1}.}
\label{fig:ocean-adv-diff}
\end{figure}

As a simple example, we consider here first the Bickley jet \cite{negrete93,rypina07},
a kinematic model for a meandering jet surrounded by vortices. We
use a quasiperiodically forced version of this velocity field, with
parameter values taken from \cite{hadjighasem16b}. Using the above
refinements to the algorithm of \cite{hadjighasem16c}, we show in
Figure \ref{fig:bickley-chi} predicted diffusion barriers for the
time interval $[0,40]$ days in the Bickley jet with quasiperiodic
time-dependence and anisotropic diffusion tensor $\mathbf{D}=\mathrm{diag}\mathrm{(2,0.5)}$.

Almost all the diffusive vortex boundaries (red), identified at time
$t=0$ as outermost closed orbits of the $\bm{\eta}_{\mathcal{T}_{0}}(\mathbf{x}_{0})$
field are larger than any of the previously detected coherent sets
in pure advection studies of this example (cf.~\cite{hadjighasem17}
and Fig.~\ref{bickley-BH-vs-diff}). In flows with non-recurrent
time dependence, invariants of the Cauchy\textendash Green strain
tensor and of its temporal average are expected to differ more, leading
to an even more significant difference between LCSs and diffusion
barriers (see Fig.~\ref{bickley-BH-vs-diff}). 
Diffusion noticeably erodes the scalar field inside closed barriers
with higher values of the transport density ${\mathcal{T}_{0}}$.
This confirms that our theory enables an a priori classification of
diffusion barriers from purely advective calculations.

The trench of the $\DBS(\mathbf{x}_{0})$ field marks the
core of the jet while ridges of the same field approximate backward-fronts
(diffusive stable manifolds). The barriers we have located indeed
prevail as organizing features of diffusive patterns, as shown in
Fig.~\ref{fig:bickley-diff} in a diffusive simulation with Péclet
number $Pe=\mathcal{O}(10^{5})$. 

Our main example, discussed in the main text involves a two-dimensional
unsteady velocity data set derived from AVISO satellite-observed sea-surface
heights (SSH) under the geostrophic approximation (cf.~\cite{haller13}
for details). As in \cite{hadjighasem16a}, our computations cover
a period of 90 days, ranging from $t_{0}=November\,11,\,2006$ to
$t_{1}=February\,9,2007$, over the longitudinal range $[-4^{\circ},6^{\circ}]$
and the latitudinal range $[-34^{\circ},-28^{\circ}]$ containing
the Agulhas leakage. This domain is covered by a regular 500x300
grid, on which we performed the steps detailed in Section 7 in the main text.

In addition to the results described in the main text, here we also show the final, evolved positions of material ring boundaries predicted solely from the satellite velocity field. Superimposed is the diffusing concentration to which the ring boundaries provide clear transport barriers (cf. Fig.~\ref{fig:ocean-adv-diff}). 

Julia and MATLAB implementations of the algorithm given in Section 7 in the main text
are available on request from the second author.
Computation times (for the Julia version) on a 2.3 GHz Intel Core i5 (DualCore)
notebook are about 50 seconds for the Bickley jet flow and about 90
seconds for the ocean flow example.
\vskip 0.5 true cm
{List of supporting animations:}

\begin{description}
\item [{SA1.mov}] Material advection  of the closed Agulhas ring boundaries, identified at time $t_0$ as outermost closed diffusion barriers. Superimposed is the diffusing concentration.

\item [{SA2.mov}] Evolution of stochastic trajectories in the Lagrangian frame, released from inside and outside the four closed diffusion barriers bounding Agulhas rings. 

\item [{SA3.mov}] Same as animation SA2.mov, but in the physical (Eulerian) frame.

\end{description}

\bibliography{pnas-main}

\begin{thebibliography}{10}

\bibitem{weiss08}
Weiss JB, Provenzale A (2008) {\em Transport and Mixing in Geophysical Flows}.
\newblock (Springer, Berlin).

\bibitem{ottino89}
Ottino J (1989) {\em The Kinematics of Mixing: Stretching, Chaos and
  Transport}.
\newblock (Cambridge University Press, Cambridge).

\bibitem{dinklage05}
Dinklage A, Klinger T, Marx G, Schweikhard L (2005) {\em Plasma Physics -
  Confinement, Transport and Collective Effects}.
\newblock (Springer, Heidelberg).

\bibitem{rosner00}
Rosner D (2000) {\em Transport Processes in Chemically Reacting Flow Systems}.
\newblock (Dover Publications).

\bibitem{toda05}
Toda M (2005) {\em Geometrical Structures of Phase Space In Multi-dimensional
  Chaos: {A}pplications To Chemical Reaction Dynamics In Complex Systems}.
\newblock (John Wiley \& Sons).

\bibitem{peacock10}
Peacock T, Dabiri J (2010) Focus issue on {L}agrangian coherent structures.
\newblock {\em Chaos} 20:017501.

\bibitem{haller15}
Haller G (2015) {L}agrangian {C}oherent {S}tructures.
\newblock {\em Annu. Rev. Fluid Mech.} 47:137--162.

\bibitem{bahsoun14}
Bahsoun W, Bose C, Froyland G (2014) {\em Ergodic Theory, Open Dynamics, and
  Coherent Structures}.
\newblock (Springer, New York).

\bibitem{peacock15}
Peacock T, Froyland G, Haller G (2015) Focus issue on the objective detection
  of coherent structures.
\newblock {\em Chaos} 25.

\bibitem{hadjighasem17}
Hadjighasem A, Farazmand M, Blazevski D, Froyland G, Haller G (2017) A critical
  comparison of {L}agrangian methods for coherent structure detection.
\newblock {\em Chaos} 27:053104.

\bibitem{press81}
Press W, Rybicki G (1981) Enhancement of passive diffusion and suppression of
  heat flux in a fluid with time-varying shear.
\newblock {\em Astrophys. J.} 248:751--766.

\bibitem{knobloch92}
Knobloch E, Merryfield W (1992) Enhancement of diffusive transport in
  oscillatory flows.
\newblock {\em Astrophys. J.} 401:196--205.

\bibitem{Thiffeault08}
Thiffeault JL (2008) Scalar decay in chaotic mixing.
\newblock {\em Lect. Notes Phys.} 744:3--35.

\bibitem{tang96}
Tang X, Boozer A (1996) Finite time {L}yapunov exponent and advection-diffusion
  equation.
\newblock {\em Physica D} 95:283--305.

\bibitem{thiffeault03}
Thiffeault JL (2003) Advection-diffusion in {L}agrangian coordinates.
\newblock {\em Phys. Lett. A} 30:415--422.

\bibitem{nakamura08}
Nakamura N (2008) Quantifying inhomogeneous, instantaneous, irreversible
  transport using passive tracer field as a coordinate.
\newblock {\em Lect. Notes Phys.} 744:137--144.

\bibitem{pratt16}
Pratt L, Barkan R, Rypina I (2016) Scalar flux kinematics.
\newblock {\em Fluids} 1:27.

\bibitem{landau66}
Landau LD, Lifshitz E (1966) {\em Fluid Mechanics}.
\newblock (Pergamon Press).

\bibitem{gurtin10}
Gurtin M, Fried E, Anand L (2010) {\em The Mechanics and Thermodynamics of
  Continua}.
\newblock (Cambridge University Press).

\bibitem{liu04}
Liu W, Haller G (2004) Strange eigenmodes and decay of variance in the mixing
  of diffusive tracers.
\newblock {\em Physica D} 188:1--39.

\bibitem{castillo08}
Castillo E, Luceno A, Pedregal P (2008) Composition functionals in calculus of
  variations. {A}pplication to products and quotients.
\newblock {\em Math. Models Methods Appl. Sci.} 18:47--75.

\bibitem{haller13}
Haller G, Beron-Vera FJ (2013) Coherent {L}agrangian vortices: the black holes
  of turbulence.
\newblock {\em J. Fluid Mech.} 731:R4.

\bibitem{serra16}
Serra M, Haller G (2016) Objective {E}ulerian coherent structures.
\newblock {\em Chaos} 26:053110.

\bibitem{haller14}
Haller G, Beron-Vera F (2014) Addendum to \textquoteleft coherent {L}agrangian
  vortices: the black holes of turbulence\textquoteright.
\newblock {\em J. Fluid Mech.} 751:R3.

\bibitem{risken84}
Risken H (1984) {\em The Fokker-Planck Equation: Methods of Solution and
  Applications}.
\newblock (Springer, New York).

\bibitem{beal11}
Beal L, De~Ruijter W, Biastoch A, Zahn R (2011) On the role of the agulhas
  system in ocean circulation and climate.
\newblock {\em Nature} 472(7344):429--436.

\bibitem{froyland15}
Froyland G, Horenkamp C, Rossi V, van Sebille E (2015) Studying an agulhas
  ring's long-term pathway and decay with finite-time coherent sets.
\newblock {\em Chaos} 25(8):083119.

\bibitem{hadjighasem16a}
Hadjighasem A, Haller G (2016) Level set formulation of two-dimensional
  {L}agrangian vortex detection methods.
\newblock {\em Chaos} 26:103102.

\bibitem{wang16}
Wang Y, Beron-Vera FJ, Olascoaga MJ (2016) The life cycle of a coherent
  lagrangian agulhas ring.
\newblock {\em J. Geophys. Res. [Oceans]} 121:3944?3954.

\bibitem{haller16}
Haller G, Hadjighasem A, Farazmand M, Huhn F (2016) Defining coherent vortices
  objectively from the vorticity.
\newblock {\em J. Fluid Mech.} 795:136--173.

\bibitem{friedman13}
Friedman A (2013) {\em Partial Differential Equations of Parabolic Type}.
\newblock (Dover Publications).

\bibitem{logan77}
Logan J (1977) {\em Invariant Variational Principles. Mathematics in Science
  and Engineering}.
\newblock Vol.{} 138, pp. 62--75.

\bibitem{moser03}
Moser J (2003) {\em Selected Chapters in the Calculus of Variations}.
\newblock (Springer, Basel).

\bibitem{lewis69}
Lewis J (1969) {\em Homogeneous functions and Euler's theorem. in: An
  Introduction to Mathematics}.
\newblock (Macmillan, London).

\bibitem{hadjighasem16c}
Hadjighasem A, Haller G (2016) Geodesic transport barriers in {J}upiter's
  atmosphere: A video-based analysis.
\newblock {\em SIAM Rev.} pp. 69--89.

\bibitem{karrasch15}
Karrasch D, Huhn F, Haller G (2015) {A}utomated detection of coherent
  {L}agrangian vortices in two-dimensional unsteady flows.
\newblock {\em Proc. R. Soc. A} 471(2173):20140639.

\bibitem{fyrillas07}
Fyrillas M, Nomura K (2007) Diffusion and {B}rownian motion in {L}agrangian
  coordinates.
\newblock {\em J. Chem. Phys.} 126:164510.

\bibitem{roessler10}
R{\"o}{\ss}ler A (2010) {R}unge--{K}utta {M}ethods for the {S}trong
  {A}pproximation of {S}olutions of {S}tochastic {D}ifferential {E}quations.
\newblock {\em SIAM J. Numer. Anal.} 48(3):922--952.

\bibitem{negrete93}
Del-Castillo-Negrete D, Morrison P (1993) Chaotic transport by rossby waves in
  shear flow.
\newblock {\em Phys. Fluids} 5:948--965.

\bibitem{rypina07}
Rypina I, et~al. (2007) On the {L}agrangian dynamics of atmospheric zonal jets
  and the impermeability of the stratospheric polar vortex.
\newblock {\em J. Atmos. Sci.} 64:3595--3610.

\bibitem{hadjighasem16b}
Hadjighasem A, Karrasch D, Teramoto H, Haller G (2016) Spectral clustering
  approach to {L}agrangian vortex detection.
\newblock {\em Phys. Rev. E} 93.

\end{thebibliography}

\end{document}